\documentstyle[12pt,epsfig]{article}
\begin{document}
%
% Here you can put your personal macros at will. E.g.:
%\def\mathrm#1{{\rm #1}}
%\def\boldmath{\rm}
\newcommand{\as}{\mbox{$\alpha_{\mathrm{s}}$}}
\newcommand{\xmu}{\mbox{$x_{\mu}$}}
\newcommand{\amz}{\mbox{$\alpha_{\mathrm{s}}(\mathrm{M_{Z^0}})$}}
\newcommand{\oaa}{\mbox{${\cal O}(\alpha_{\mathrm{s}}^2$})}
\newcommand{\epem}{\mbox{$\mathrm{e^+e^-}$}}
\newcommand{\Zzero}{\mbox{${\mathrm{Z}^0}$}}
\newcommand{\WW}{\mbox{$\mathrm{W^+W^-}$}}
\newcommand{\ZZ}{\mbox{$\mathrm{ZZ}$}}
\newcommand{\qq}{\mbox{$\mathrm{q\overline{q}}$}}
\newcommand{\ppbar}{\mbox{$\mathrm{p\overline{p}}$}}
\newcommand{\Zqq}{\mbox{$ \Zzero / \gamma \rightarrow \qq $}}
\newcommand{\WWqqqq}{\mbox{\WW$\rightarrow$\qq\qq}}
\newcommand{\ZZqqqq}{\mbox{\ZZ$\rightarrow$\qq\qq}}
\newcommand{\WWqqln}{\mbox{\WW$\rightarrow$\qq$\ell\overline{\nu_{\ell}}$}}
\newcommand{\mz}{\mbox{$M_{\mathrm{Z}^0}$}}
\newcommand{\mw}{\mbox{$M_{\mathrm{W}}$}}
\newcommand{\ecm}{\mbox{$E_{cm}$}}
\newcommand{\Opal}{\mbox{O{\sc pal}}}
\newcommand{\Aleph}{\mbox{A{\sc leph}}}
\newcommand{\Delphi}{\mbox{D{\sc elphi}}}
\newcommand{\LepII}{\mbox{LEP2}}
\newcommand{\LepI}{\mbox{LEP1}}
\newcommand{\Jetset}{\mbox{J{\sc etset}}}
\newcommand{\Pythia}{\mbox{P{\sc ythia}}}
\newcommand{\Herwig}{\mbox{H{\sc erwig}}}
\newcommand{\half}{\mbox{$\textstyle\frac{1}{2}$}}
\newcommand{\boldp}{\mbox{\boldmath$p$}}
\newcommand{\boldn}{\mbox{\boldmath$n$}}
\newcommand{\Mbar}{\mbox{$\overline{M}$}}
\newcommand{\costh}{\mbox{$\cos\theta$}}
\newcommand\beq{\begin{equation}}
\newcommand\eeq{\end{equation}}
%
%%%%%%%%%%%%%%%%%%%%%%%%%%%%%%%%%%%%%%%%%%%%%%%%%%%%%%%%%%%%%%%%%%%%%%%%%%%%%%%
%
% local definitions
% -----------------
%
\newcommand\new{\newcommand}         % shorthand for \newcommand
\newcommand\ren{\renewcommand}       % shorthand for \renewcommand
\new\mbf{\mathbf}
\new\mrm{\mathrm}                    % shorthand for \mathrm
\new\mm[1]{{\mbox{\hspace{#1mm}}}}   % create horizontal space
\new\ee{\mbox{$e^+e^-$}}
\new\EE{\mbox{$\mbf{e^+e^-}$}}
\new\asmz{\mbox{$\alpha_s(M_Z)$}}
\new\MSbar{\mbox{$\overline{\mrm{MS}}$}}
\new\asmuF{\mbox{$\alpha_s(\mu_F)$}}
\new\asmuR{\mbox{$\alpha_s(\mu_R)$}}
\def\cO#1{{{\cal{O}}}\left(#1\right)}
\def\lrang#1{\left\langle #1 \right\rangle}
\def\beql#1{\begin{equation}\label{#1}}   \def\eeq{\end{equation}}
\newcommand{\etal}{{\sl et al.}}
\def\pl#1#2#3{{\it Phys. Lett. }{\bf #1}(19#2)#3}
\def\zp#1#2#3{{\it Z. Phys. }{\bf #1}(19#2)#3}
\def\prl#1#2#3{{\it Phys. Rev. Lett. }{\bf #1}(19#2)#3}
\def\rmp#1#2#3{{\it Rev. Mod. Phys. }{\bf#1}(19#2)#3}
\def\prep#1#2#3{{\it Phys. Rep. }{\bf #1}(19#2)#3}
\def\pr#1#2#3{{\it Phys. Rev. }{\bf #1}(19#2)#3}
\def\np#1#2#3{{\it Nucl. Phys. }{\bf #1}(19#2)#3}
\def\sjnp#1#2#3{{\it Sov. J. Nucl. Phys. }{\bf #1}(19#2)#3}
\def\app#1#2#3{{\it Acta Phys. Polon. }{\bf #1}(19#2)#3}
\def\ar#1#2#3{{\it Ann.\ Rev.\ Nucl.\ Part.\ Sci.\ } {\bf #1} (19#3) #2}
\def\cav#1{Cambridge preprint Cavendish--HEP--#1}
\def\eV{{\rm e\kern-0.12em V}}  \def\TeV{{\rm T}\eV}
\catcode `@ 11
\def\biblabel#1{\if@filesw\immediate
\write\@auxout{\string\bibcite{#1}{\the\value{\@listctr }}}\fi}
\catcode `@ 12
\noindent hep-ph/9602288\newline
\begin{center}{\large \bf QCD\footnote{To appear on
the Report of the Workshop on Physics at LEP2, CERN 96-01, vol.~1, 1996.}
}\end{center}
\begin{center}
{\it conveners}:  P. Nason and B.R. Webber
\end{center}
\begin{center}
{\it Working group}:
D.Ward,  D. Lanske, L.A. del Pozo, F. Fabbri and B. Poli (OPAL),
G.Cowan and C.Padilla (ALEPH), M. Seymour, F. Hautmann,
Yu.L. Dokshitzer and V.A. Khoze.
\end{center}
\vspace*{1.0cm}
{\def\footnote#1{} \tableofcontents}
\newpage
\section{Introduction}% \hfill {\normalsize\it P. Nason and B.R. Webber}
\begin{figure}[hpbt]
\centerline{
\epsfig{figure=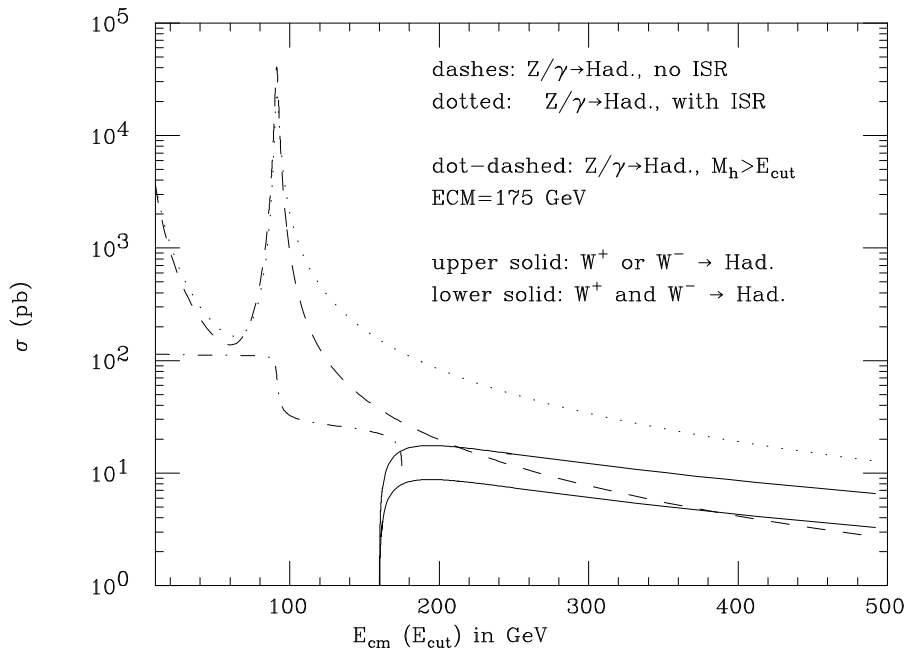,width=12cm}
}
\caption{
Hadronic cross sections as a function of the centre of mass energy.}
\label{sigmaee}
\end{figure}

\label{QCDintro}
LEP1 has performed a gigantic
task in testing QCD predictions.
In this it has benefited from the very large statistics available, the
substantial lack of background, and the fact that initial state radiation
plays only a minor r\^ole on the resonance.
At LEP2, QCD tests are more challenging. Initial state radiation is
very important, there is a $WW$ production background, and statistics
are somewhat limited.
In fig.~\ref{sigmaee}
we show the annihilation cross section as a function of the centre of mass
energy.
The figure reports the Born cross section for the production of hadronic final
states through the $Z/\gamma$ annihilation process, and the same cross section
with the inclusion of the initial state radiation. This
increases the cross section considerably
above the $Z$ resonance due to the $e^+e^-\to Z\,\gamma$
process, in which the hadronic system has an invariant mass equal to the mass
of the $Z$ boson. In the figure we also show the hadronic cross section
at a fixed $E_{\rm cm}$, as a function of a lower cut $E_{\rm cut}$
on the invariant mass
of the hadronic system (dot--dashed line). With $E_{\rm cut}= 0$ this cross
section coincides with the value of the dotted line at 175 GeV. As the cut is
increased above the $Z$ mass, the cross section drops suddenly, and it
approaches the partonic cross section at 175 GeV. As the cut approaches 175
GeV the cross section vanishes, but it is quite clear that if we allow
for few GeV of initial state radiation, its value is very close
to the Born cross section. Assuming therefore a 20 pb cross section,
with an integrated luminosity of 500 pb$^{-1}$
we expect 10000 hadronic events. From the figure we see
that the $W$ background is not a negligible one, and further cuts
should be imposed to get rid of it. From statistics alone, the error on
a measurement of the total hadronic cross section is 1\%. Since
$\sigma_{\rm tot}=\sigma^{(0)}_{\rm tot}(1+\as/\pi+\ldots)$, we would expect
a 25--30\% error on a determination of $\as$ from the hadronic
cross section,
not including systematics. Therefore, a useful measurement of $\as$ from the
total cross section will not be possible at LEP2.
Instead, it will
be possible to determine $\as$ from jets.
The rule of thumb in these cases is that we expect most events to be two--jet
events, a fraction $\as$ of three--jet events, and a fraction $\as^2$
of 4--jet events. With 10000 hadronic events, we would have 1000 three--jet
events, which will allow us to determine $\as$ with a statistical precision
of 3\%. Assuming that $\as(M_Z)=0.123$, we expect $\as(175\;{\rm GeV})=0.112$,
a 10\% variation. It seems therefore possible to see the running of $\as$
between LEP1 and LEP2.

A large fraction of this report will be dedicated to the problem of measuring
$\as$ from jets at LEP2. In Section~\ref{Ward} the relevant experimental
aspects of event selection and background corrections will be dealt with.
In the Sections~\ref{Nason} and \ref{Seymour} the present status
of theoretical calculations for jet shape variables
will also be given.

Using the large number of hadronic events, studies of particle spectra
will certainly be possible. Section~\ref{Padilla}
is dedicated to fragmentation
function studies at LEP2. The study of fragmentation functions is
a relatively recent topic at LEP1. Measurements
of the various components of the quark and gluon fragmentation functions
have been performed at LEP1, and they allow us to
make an absolute prediction for the fragmentation function
at LEP2 energies, and also for the fragmentation function in W decays.
We will see that it is very difficult to see scaling violation effects
from LEP1 to LEP2. It is nevertheless important to measure the
fragmentation function to check for the consistency of the whole approach,
since important assumptions are often made when performing the fit
(for example,
flavour SU(3) symmetry).
A study of scaling violation towards the small $x$ region
has not yet been performed even at LEP1, mostly because of the lack of
a complete theoretical calculation. We will present
the relevant theoretical ideas
in Section~\ref{Hautman}.

Section~\ref{Fabbri} will be dedicated to the measurement of particle
multiplicities
at LEP2. QCD makes a prediction for the energy dependence of the multiplicity,
and for the shape of the multiplicity distribution,
based upon the assumption known as local parton--hadron duality.
The measurement of the multiplicity in heavy--flavoured events has recently
received some attention, and will also be considered here.
Based again upon the idea of local
parton--hadron duality, QCD predicts
many features of the small--$x$ particle spectrum and correlations.
Section~\ref{DelPozo} will deal with these topics.

\section{Event Selection and Event
Shapes -- Experimental\protect\footnote{The present Section is mostly
work of D. Ward, including contributions from S.Bethke, G.Cowan,
D. Lanske, and C.Padilla.}}
\label{D_Ward}\label{Ward}
%{\it D.Ward,  D. Lanske, S.Bethke, G.Cowan and C.Padilla}\newline
\subsection{Introduction}
A number of interesting studies of QCD may be performed at \LepII\
using \Zqq\ events.  Although the number of events will be much smaller
than at \LepI, it may be sufficient to explore aspects of the energy
evolution of QCD.  In this study we focus on the determination of
\as.  The value of \amz\ has been determined using
a number of techniques involving jet rates and event shape 
observables at \LepI
\cite{A-as1,A-as2,D-as1,D-as2,L-as,O-grandas,O-asnlla,O-colfac}
and at SLD~\cite{jp_SLD}. For example, using
a combination of resummed next-to-leading log (NLLA) and \oaa\ QCD 
calculations, an average measurement of 
\[ \amz\ = 0.123 \pm 0.006 \]
was obtained~\cite{bib-Siggi}.  
Taking the typical centre-of-mass energy at \LepII\ to be
175~GeV, we may expect the value of \as\ to be reduced to 0.112.
Although the change in \as\ is not great compared to the uncertainty on 
the \LepI\ measurement, it should be noted that the error at \LepI\
is predominantly theoretical in origin, and thus may be largely correlated
between \LepI\ and \LepII.  We may therefore hope to make a useful measurement
of the difference in \as\ between the two energies.

The experimental difficulties at \LepII\ are somewhat different from
those at \LepI.  At \LepI\, hadronic \Zzero\ decays could be readily 
identified with efficiencies in excess of 98\%, and with negligible background.
At \LepII\ there are extremely large radiative corrections, and W$^+$W$^-$
events may contribute a significant and troublesome background.  
Therefore, in Sect.~\ref{sect-Zsel} we investigate the problems of selecting 
a sample of non-radiative \Zqq\ events, and discuss the extent to which 
these selection procedures may bias the events selected.

It will turn out that the events which may be selected most cleanly 
are those nearer to the 
two-jet region.  Multi-jet events are much more susceptible to contamination
from  W$^+$W$^-$ events.  Since statistics are also meagre, 
and most of the events lie in the two jet region, this suggests
that techniques based on the resummed NLLA QCD calculations will be most
effective in determining \as, since these calculations are expected
to describe the two-jet region best.    We have therefore focused on 
those event shape variables for which complete resummed NLLA
calculations are available,
namely Thrust ($T$), heavy jet mass ($M_H$), total jet broadening
($B_T$) and wide jet broadening ($B_W$)~\cite{bib-catani}.  
We also examine jet rates
in the Durham jet-finding scheme, for which NLLA calculations
are available -- specifically the observable
$y_{23}^{(D)}$ which is the value of $y_{cut}$ at which the event 
changes from two- to three-jet.  
All these variables are discussed in, for example, Ref.~\cite{O-asnlla}.
The next-to-leading order calculation~\cite{bib-catani2}
for $y_{23}^{(D)}$ as used so far 
by the LEP experiments was known to be incomplete.  Recently, however,
a more complete calculation has been presented~\cite{bib-schmelling}.
We have not yet studied this new calculation, for compatibility with the 
existing \LepI\ results. 
Resummed calculations are also now available for 
the C-parameter~\cite{bib-webber}, 
though they are not yet published, and are therefore not discussed here.

\subsection{Selection of \mbox{\boldmath \Zqq} events}
\label{sect-Zsel}
The discussion here will be based on events generated with 
\Pythia~\cite{bib-pythia} version 5.715, with hadronization parameters
tuned to \LepI\ data~\cite{O-jttune}.  
The examples given below will relate to events
processed through the \Opal\ detector simulation, but it is to be expected 
that similar results would hold for the other experiments.
The cross-sections predicted 
for \Zqq\ events and for the principal source of background
\WWqqqq\ are as follows:
\begin{center}
\begin{tabular}{|l|r|r|r|}
\hline
         & \multicolumn{3}{c|}{ Cross-section / pb}\\
Reaction & 161~GeV & 175~GeV & 192~GeV \\
\hline
$\epem\rightarrow\WW\rightarrow\qq\qq$  &  1.69 & 6.34 & 7.92 \\
\hline
$\epem\rightarrow\Zzero/\gamma\rightarrow\qq$  & 149.6 & 116.8 & 90.6  \\
$\epem\rightarrow\Zzero/\gamma\rightarrow\qq$ ; $E_{isr}<$ 30~GeV 
& 39.0 & 29.9 & 22.4  \\
$\epem\rightarrow\Zzero/\gamma\rightarrow\qq$ ; $E_{isr}<$ 1~GeV 
& 26.1 & 20.0 & 15.1  \\
\hline
\end{tabular}
\end{center}
The \Zqq\ cross-section is also given for two cuts on the amount of energy
lost in initial state radiation.  The useful cross-section for QCD 
studies is the non-radiative cross-section.  The cut at 30~GeV 
corresponds roughly to the minimum in the hadronic mass spectrum
$d\sigma/d M_h$ between the non-radiative 
process and radiation down to the \Zzero\ pole.  Unless otherwise stated,
the results shown relate to 175~GeV.  

It is helpful to consider the selection of \Zqq\ events in two stages.
In stage I we remove the leptonic and highly radiative events, and the
\WWqqln\ events, mainly using cuts on multiplicity and energy/momentum 
balance.  These cuts introduce rather little bias into the \Zqq\ 
event sample.  The stage II cuts are to remove \WWqqqq\ events, and 
are more problematic, since they turn out to bias the selected 
\Zqq\ sample significantly.

Typical stage I cuts would be as follows:
\begin{itemize}
\item 
Require $ |\cos\theta_T|<0.9 $
to ensure reasonable containment of the event, 
where $\theta_T$ is the polar angle of the thrust axis.
\item 
Require the number of charged tracks to be $ \mathrm{N}_{ch}>6 $ 
to remove purely leptonic
events.  This cut causes a negligible loss of \Zqq\ events.
\item
In Fig.~\ref{fig-drw1} 
we plot $R_{vis}$ against $R_{miss}$ for various classes 
of events, where $R_{vis}$ is the visible energy scaled by the centre of mass
energy $E_{c.m.}$, and $R_{miss}$ is the missing momentum scaled by 
$E_{c.m.}$.
%%%%%%%%%%%%%%%%%%%%%%%%%%%%%%%%%%%%%%%%%%%%%%%%%%%%% 
\begin{figure}[hbtp]
\centerline{
\epsfig{figure=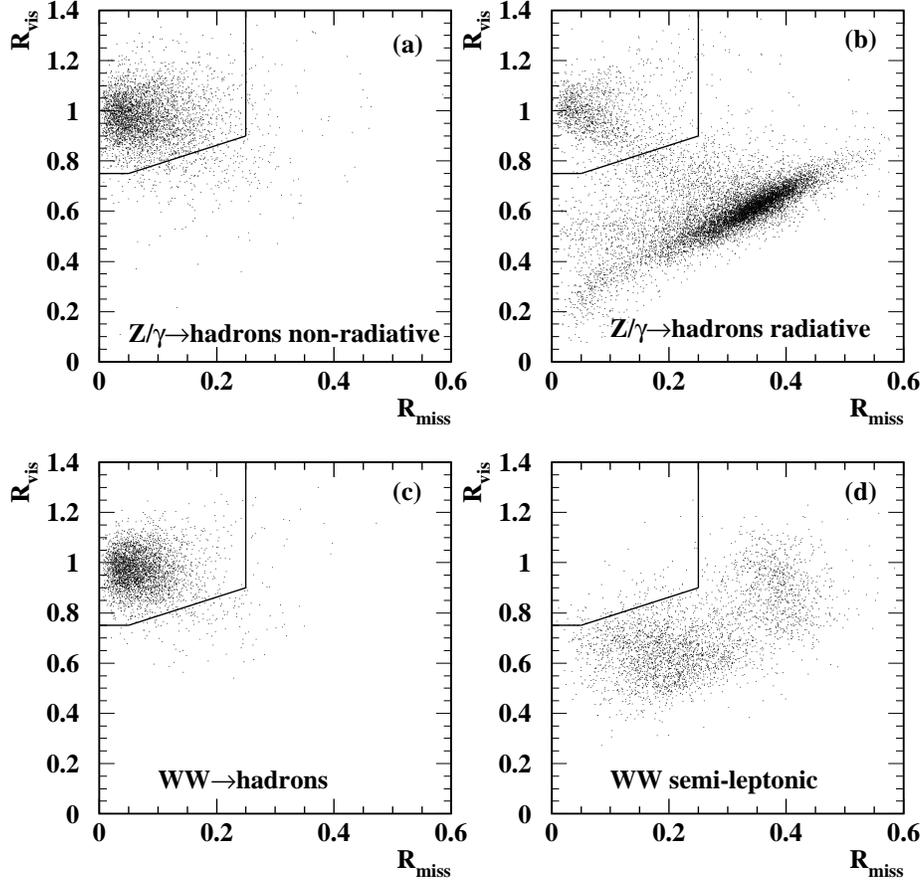,width=12cm}
%\epsfysize=12cm
%\epsffile[20 150 550 650]{drw1.ps}
}
\caption{Plots of $R_{vis}$ against $R_{miss}$ for 
(a) \Zqq\ events having less than 30~GeV initial state radiation
(b) \Zqq\ events having more than 30~GeV initial state radiation
(c) \WWqqqq\ events
(d) \WWqqln\ events.
The lines show typical cuts.  These plots are at $E_{cm}=175$~GeV, 
though they are only weakly energy dependent.}
\label{fig-drw1}
\end{figure}
%%%%%%%%%%%%%%%%%%%%%%%%%%%%%%%%%%%%%%%%%%%%%%%%%%%%% 
It is desirable to have the best resolution on 
the visible energy and missing momentum, which involves using an algorithm 
to combine the information from the charged tracks, electromagnetic and 
hadronic calorimeters so as to reduce double counting.  
We note that the non-radiative \Zqq\ events and the \WWqqqq\ events 
are peaked around $R_{vis}=1$ and $R_{miss}=0$.  The \WWqqln\
events, and most of the radiative \Zqq\ events lie away from this point.
Typical  cuts are  shown by the lines in Fig.~\ref{fig-drw1}.
\item
Fig.~\ref{fig-drw1}(b) reveals a group of radiative \Zqq\ events having
$R_{vis}\sim 1$ and $R_{miss}\sim 0$.
In these events, the radiative photons are detected in the electromagnetic 
calorimeter.  Such photons may be identified using standard criteria 
on lateral shower shapes.
The cluster should also be required to be be isolated, 
for example by demanding that within a cone of half angle 0.2~rad
centred about the cluster less than 1~GeV is observed.
If the energy of the most energetic cluster satisfying the above criteria
exceeds \mbox{$0.6 \times p_{\gamma}$}, the event is rejected.  Here, 
$p_{\gamma}$ is the expected photon momentum in an 
\mbox{$\epem \rightarrow \Zzero \gamma$} event, 
i.e. \mbox{$p_{\gamma}=(E_{c.m.}^2-M_Z^2)/2E_{c.m.}$}.
\end{itemize}
The cross-sections for the various channels of interest before and after 
these stage I selection cuts are listed in table~\ref{tab-drw1}.
\begin{table}[htb]
\begin{center}
\begin{tabular}{|c|c|c|}
\hline
Channel & Cross-section /pb & Cross-section /pb \\
        & $|\cos\theta_T|<0.9$ & after stage I cuts \\
\hline
\Zqq\ ($E_{isr}<1$~GeV)  &   17.90   &  16.51 \\
\hline
\Zqq\ ($E_{isr}<30$~GeV) &   26.68   &  23.99 \\
\hline
\Zqq\ ($E_{isr}>30$~GeV) &   73.37   &   1.07 \\
\hline
\WWqqln\                 &    5.88   &   0.06 \\
\hline
\WWqqqq\                 &    6.08   &   5.74 \\
\hline
\end{tabular}
\caption{Cross-sections at $E_{cm}=175$~GeV, based on \Pythia.}
\label{tab-drw1}
\end{center}
\end{table}
Hence, at $E_{cm}=175$~GeV, 
in the region $|\cos\theta_T|<0.9$, the stage~I cuts accept
92\% of the non-radiative \Zqq\ events, whilst accepting only 
around 1.5\% of the radiative \Zqq\ events and \WWqqln\ events.
The \WWqqqq\ events are accepted with high efficiency. 
The corresponding figures at 192~GeV and 161~GeV are essentially the same.
Backgrounds from two-photon
events, Z\epem\ and W$e\nu$ final states have been examined, and 
appear to be negligible.
\ZZqqqq\ does contribute, but at a much lower rate than 
\WWqqqq, with similar characteristics.

The main feature which distinguishes the \WWqqqq\ events
(and also the much smaller contribution from \ZZqqqq) from the \Zqq\
events is that the former contain four quarks, and thus generally have  
four or more jets, and are therefore less collimated.  
Furthermore, the invariant masses
of appropriate pairs of jets should equal the mass of the W boson.
We have examined the use of the following variables in separating these
event classes:
\begin{itemize}
\item
The ``narrow jet broadening'', $B_N$.  
The event is divided into two hemispheres, $S_{\pm}$, 
by the plane orthogonal to the thrust axis, $\hat{\boldn}_T$. 
In each hemisphere, the quantity
$ B_{\pm} = \sum_{i\in S_{\pm}}|{\boldp}_i \times \hat{\boldn}_T| /
      2\sum_i |{\boldp}_i| $
is computed, where the sum in the denominator runs over all particles, 
whilst that in the numerator runs over one hemisphere.  
$B_N$ is defined by
$B_N = \min(B_+,B_-) $.
\item
The scaled ``light hemisphere mass'', $M_L/E_{vis}$.  
The event is divided into two hemispheres, $S_{\pm}$, 
by the plane orthogonal to the thrust axis, and the invariant mass
of each is computed, $M_{\pm}$.   
Then, $M_L$ is defined by $M_L = \min(M_+,M_-) $.
\item
The value of $y_{cut}$ at which the event changes from 3-jet to 
4-jet in the Durham jet finding scheme, $y_{34}^{(D)}$.  
\item 
Using the Durham jet finder, the event may be forcibly reconstructed as 
having four jets.  The invariant masses of pairs of jets may be formed, 
from which we define the variable 
$ D^2 = \min \left[ (M_{ij}-M_W)^2+(M_{kl}-M_W)^2 \right] $
where the minimum is taken over the permutations $(ij;kl)=(12;34), (13;24),
(14;23)$. 
Various ways of scaling the jet energies in order to improve 
the W mass resolution have been considered in connection with the 
W mass determination, but have not been used here. 
\end{itemize}

\begin{figure}[hbtp]
\vspace{0.1cm}
\centerline{
\epsfig{figure=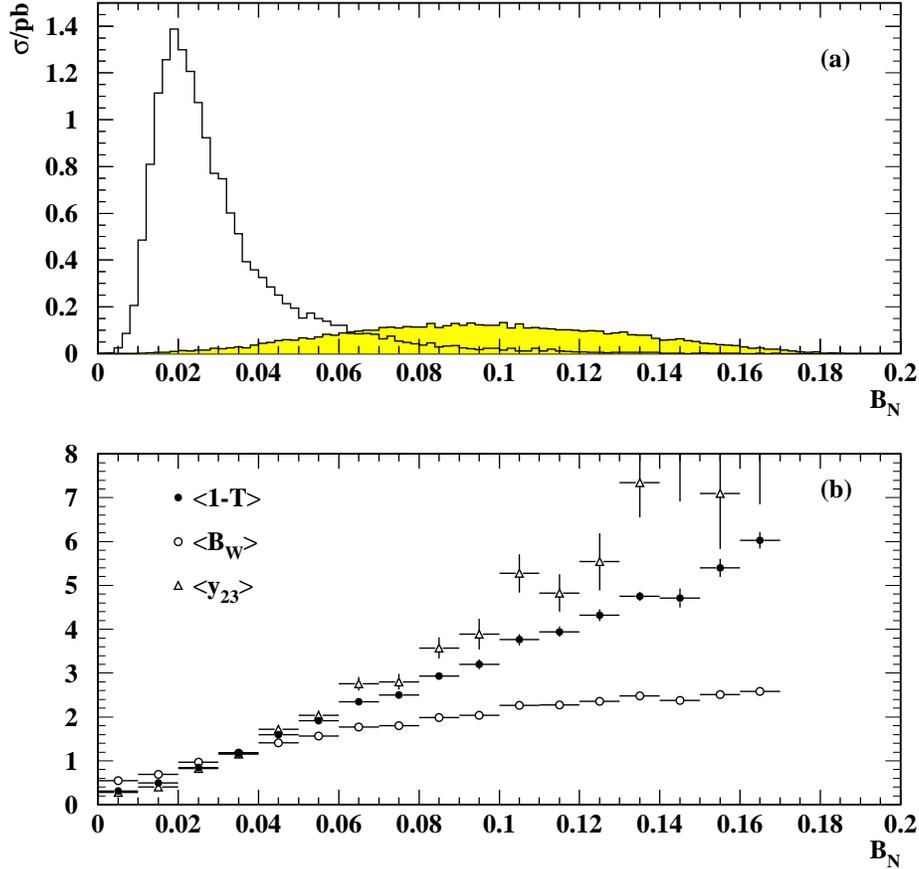,width=12cm}
%\epsfysize=12cm
%\epsffile[20 150 550 650]{drw2.ps}
}
\caption{(a) Distributions (at 175~GeV) of $B_N$ for \Zqq\ events having less 
than 1~GeV initial state radiation (open histogram) and for 
\WWqqqq\ events (shaded)
(b) average values of $(1-T)$, $B_W$ and $y_{23}^{(D)}$ (scaled by their
overall mean values) as a function of $B_N$}
\label{fig-drw2}
\end{figure}

In Fig.~\ref{fig-drw2}(a) we show the distributions of $B_N$ 
for non-radiative \Zqq\ events ($E_{isr}<1$~GeV) and for \WWqqqq\ events,
after the stage I cuts.
%%%%%%%%%%%%%%%%%%%%%%%%%%%%%%%%%%%%%%%%%%%%%%%%%%%%% 
%%%%%%%%%%%%%%%%%%%%%%%%%%%%%%%%%%%%%%%%%%%%%%%%%%%%% 
In order to judge the correlation between
$B_N$ and the observables which we would wish to use for the determination 
of \as\, we show in Fig.~\ref{fig-drw2}(b) the average values of 
$(1-T)$, $B_W$ and $y_{23}^{(D)}$ (normalized to their overall mean
values)
for non-radiative \Zqq\ events as a function of $B_N$.
It is evident that the \WWqqqq\ contribution can be reduced to almost any
level desired by cutting on $B_N$, but at an increasing cost in bias, and
a corresponding loss in statistics.
Generally, $M_L$ and  $y_{34}^{(D)}$ show similar behaviour to $B_N$.
The $D^2$ variable offers a less clean separation between 
the \Zqq\ and \WWqqqq\ events, but it appears that it may introduce somewhat 
less, or different,  bias, and may thus be complementary.

These observations may be quantified in table~\ref{tab-drw2} below, 
where we show the effect of various possible stage II cuts on the
\Zqq\ non-radiative signal and the \WWqqqq\ background.\\
We give the average values of $1-T$, $B_W$ and $y_{23}^{(D)}$
as an indication of the bias caused by the cuts.
\begin{table}[htbp]
\begin{center}
\begin{tabular}{|l|c|c|c|c||c|}
\hline
& \multicolumn{4}{c||}{\Zqq ($E_{isr}<1$~GeV)} & \WWqqqq \\
\cline{2-6}
Cut(s)       & $\sigma$~/pb & $<1-T>$ & $<B_W>$ & $<y_{23}^{(D)}>$ & 
               $\sigma$~/pb \\
\hline
\hline
$|\costh_{T}|<0.9$ & 17.90 & 0.0598 & 0.0708 & 0.0196 & 6.08 \\
\hline
Stage I only       & 16.51 & 0.0587 & 0.0701 & 0.0195 & 5.74 \\
\hline
$B_N<$0.07         & 15.73 & 0.0528 & 0.0668 & 0.0169 & 1.34 \\
$B_N<$0.06         & 15.27 & 0.0504 & 0.0653 & 0.0160 & 0.87 \\
$B_N<$0.05         & 14.53 & 0.0474 & 0.0632 & 0.0149 & 0.50 \\
$B_N<$0.04         & 13.26 & 0.0433 & 0.0601 & 0.0133 & 0.24 \\
\hline
$M_L<$0.175        & 15.26 & 0.0504 & 0.0656 & 0.0161 & 1.37 \\
\hline
$y_{34}^{(D)}<$0.0065 & 15.34 & 0.0501 & 0.0644 & 0.0156 & 0.88 \\
\hline
$D^2>$300~GeV$^2$  & 14.17 & 0.0543 & 0.0667 & 0.0164 & 1.90 \\
$D^2>$600~GeV$^2$  & 12.25 & 0.0504 & 0.0638 & 0.0140 & 0.89 \\
\hline
$D^2>$600~GeV$^2$ {\em and}  $B_N<$0.06 &
                     11.71 & 0.0460 & 0.0613 & 0.0124 & 0.24 \\
$D^2>$300~GeV$^2$ {\em and}  $B_N<$0.05 &
                     12.79 & 0.0457 & 0.0615 & 0.0132 & 0.24 \\
$B_N-\sqrt{D^2}/2000<0.03$  & 
                     13.84 & 0.0451 & 0.0612 & 0.0137 & 0.25 \\
\hline
\end{tabular}
\caption{Cross-sections at 175~GeV accepted after the Stage I cuts, 
and after various possible Stage II cuts. 
The average values of various relevant observables are also shown,
to indicate the level of bias introduced. }
\label{tab-drw2}
\end{center}
\end{table}
We note that the stage I cuts cause only a small bias.
We show several possible cuts on the $B_N$ variable.  
The background from \WWqqqq\ may be reduced, for example, to a level of 4\%
with an efficiency for selection \Zqq\ events of 82\%.  However, the
sample of \Zqq\ events accepted is strongly biased.  The bias, as 
measured by the mean value of the observable, tends to be 
greatest for $y_{23}^{(D)}$ and smallest for $B_W$.  
We show similar results for cuts on $M_L$ and $y_{34}^{(D)}$, 
where we have chosen cuts which yield roughly 
the same \Zqq\ efficiency as the $B_N<0.06$ cut.
Cutting on $M_L$ is less effective than $B_N$
at removing \WWqqqq\ background, 
while a cut on $y_{34}^{(D)}$ gives essentially the same performance as
$B_N$.
The cut on $D^2>600$~GeV$^2$ yields the same \WWqqqq\ contamination 
(7\%) as the $B_N<0.06$ cut, but for a significantly lower 
\Zqq\ efficiency (69\% compared to 85\%).  
Using $D^2$ yields a somewhat smaller bias on $1-T$,
but the bias on $y_{23}^{(D)}$ is a little greater.
The two observables $B_N$ and $D^2$ are not strongly 
correlated (whereas, for example, $B_N$ and $M_L$ are 
highly correlated), suggesting that a joint cut on the two 
variables could give better separation.  Examples are given in 
Table~\ref{tab-drw2}.  The \WWqqqq\ background may, for example, 
be reduced to around the 2\% level for a \Zqq\ efficiency of 
almost 80\%, with somewhat less bias than a cut on $B_N$ alone.  
The precise cuts chosen for the separation of 
\Zqq\ and \WWqqqq\ events may therefore need to depend on the analysis being 
performed -- whether a high purity is demanded, or whether a 
comparatively unbiased sample is required.

\begin{figure}[hbtp]
\centerline{
\epsfig{figure=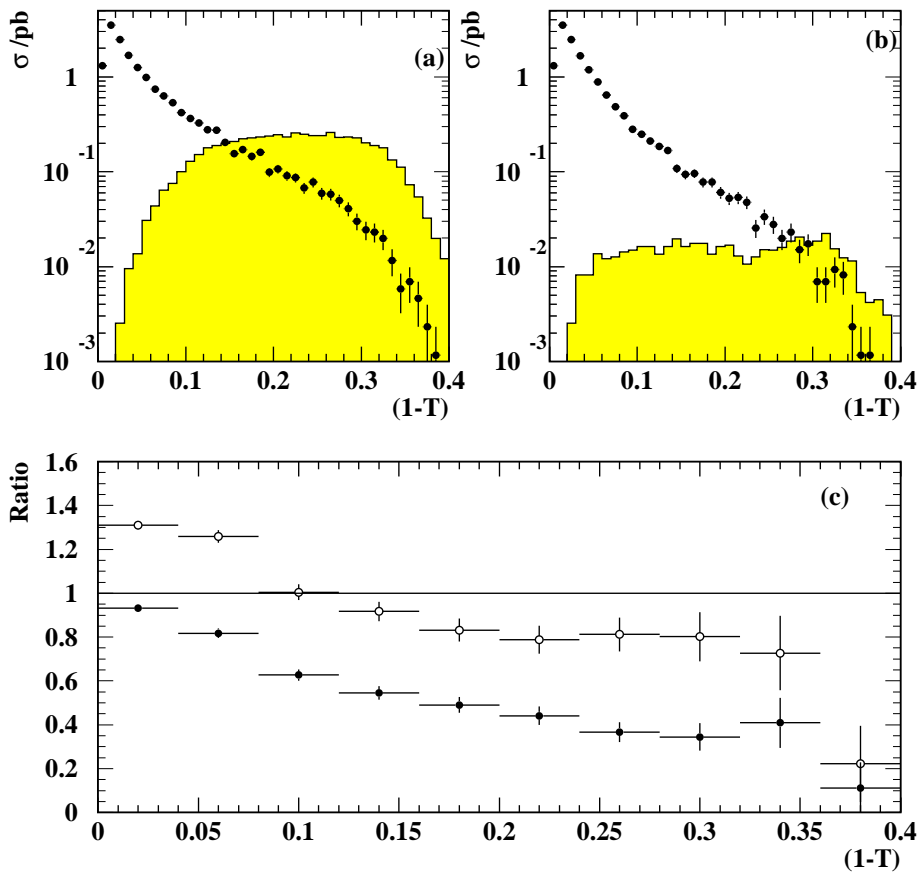,width=12cm}
%\epsfysize=12cm
%\epsffile[20 150 550 650]{drw3.ps}
}
\caption{(a) Distributions (at 175~GeV) of $(1-T)$ after the stage I cuts.  
\Zqq\ non-radiative ($E_{isr}<1$~GeV) events are shown as points with errors, 
and \WWqqqq\ events by the shaded histogram.
(b) as (a), after applying the stage I cuts and the stage II cut $B_N<0.05$.
(c) Biases to the distribution of $(1-T)$.
The stage I cuts and the stage II cut $B_N<0.05$
are applied.  The closed points show the fraction of 
\Zqq\ non-radiative ($E_{isr}<1$~GeV) events accepted after cuts.
The open points show the ratio of all accepted \Zqq\ events after cuts 
to non-radiative ($E_{isr}<1$~GeV) \Zqq\ events before cuts.}
\label{fig-drw3}
\end{figure}

In Fig.~\ref{fig-drw3}(a) we show the distributions of a typical
observable which  
may be used for the determination of \as, $(1-T)$,
after the stage I cuts.
We compare the \Zqq\ non-radiative ($E_{isr}<1$~GeV) signal
with the \WWqqqq\ background.  
In Fig.~\ref{fig-drw3}(b) we show the same distributions after the 
stage II cuts, taking $B_N<0.05$ as a typical stage II cut.
%%%%%%%%%%%%%%%%%%%%%%%%%%%%%%%%%%%%%%%%%%%%%%%%%%%%% 
%%%%%%%%%%%%%%%%%%%%%%%%%%%%%%%%%%%%%%%%%%%%%%%%%%%%% 
As expected, the background tends to be concentrated toward large values
of $(1-T)$, i.e. the region of hard gluon emission in the
\Zqq\ reaction.  The two-jet region of the \Zqq\ process is relatively free of
background.  Other stage II cuts give similar results.
In Fig.~\ref{fig-drw3}(c) we show the efficiency of the stage I+II cuts, 
taking $B_N<0.05$ as a typical stage II cut, as a function of the
$(1-T)$, for \Zqq\ non-radiative ($E_{isr}<1$~GeV)
events (solid points).   As expected, the cuts bias against large
values of $(1-T)$.  
In Fig.~\ref{fig-drw3}(c) we also show as open points the ratio of 
the distributions of all accepted \Zqq\ events (including radiative
events) to those of the non-radiative events before selection cuts.
In general the effect of initial state radiation is to bias the
distribution towards higher values, but this is counteracted by the
tendency of the cuts to reject events with high values of the
observables.  The net effect is that the distribution of the 
accepted radiative \Zqq\ events is quite similar to the 
distribution of non-radiative events before cuts, and so the 
ratios in Fig.~\ref{fig-drw3}(c) are increased roughly uniformly.
Other stage II cuts give similar results, though the efficiencies may be 
systematically higher or lower.

\subsection{Determination of \mbox{\boldmath \as } }
\label{sect-as}

\begin{figure}[hbtp]
\centerline{
\epsfig{figure=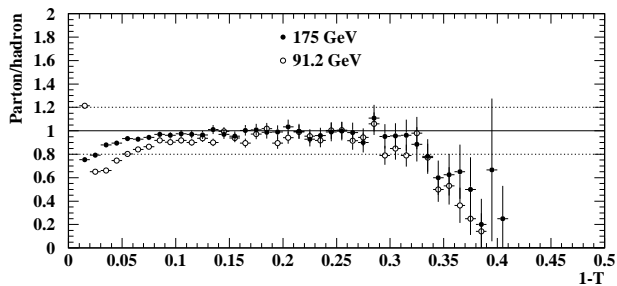,width=8cm}
%\epsfysize=8cm
%\epsffile[20 150 550 450]{drw4.ps}
}
\caption{Hadronization corrections for the distributions of $(1-T)$.}
\label{fig-drw4}
\end{figure}

Before comparing with QCD calculations, 
the observed data must be corrected for 
the effects of detector resolution, the acceptance
of selection cuts and the effects of background
(Fig.~\ref{fig-drw3}).  The influence of hadronization must then be 
accounted for, and one standard way of doing this is to multiply the 
corrected hadron level data by the ratio of the parton level to hadron 
level distributions from a Monte Carlo model.  In Fig.~\ref{fig-drw4}
we show these ratios for $(1-T)$, based on \Jetset7.4, 
at 175~GeV (\LepII) and 91.2~GeV (\LepI).
%%%%%%%%%%%%%%%%%%%%%%%%%%%%%%%%%%%%%%%%%%%%%%%%%%%%% 
%%%%%%%%%%%%%%%%%%%%%%%%%%%%%%%%%%%%%%%%%%%%%%%%%%%%% 
We note that the correction factors
at \LepII\ are significantly closer to unity, especially at small values
of $(1-T)$, corresponding to the two-jet region.  Similar comments apply to
the other observables.    
A requirement for a credible analysis is that the correction factors be not 
too far from unity.  

In this study, we investigate three types of QCD calculations, which
may be used as the basis of a measurement of \as\ from event shape
variables.  These are:
\begin{description}
\item[\oaa] 
The QCD matrix elements, expanded as a power series in \as\ are fully
known to \oaa~\cite{bib-ert}.  From previous studies at \LepI\
we know that these calculations are applicable 
in the ``3-jet'' region, i.e. the region dominated by hard gluon radiation.
A significant uncertainty in applying the \oaa\ calculations is the
choice of renormalization scale, $\mu$, represented by $\xmu=\mu/E_{c.m.}$.
The region over which the data can successfully be fitted can be 
extended further into the 2-jet region by choosing a small value of 
$\xmu \sim 0.1$ (typically).
\item[NLLA] 
In the 2-jet region, the expansion in powers of \as\ is bound to fail, 
because large logarithms arise associated with collinear and soft gluon 
emission.  In this region, "NLLA" calculations are available which resum 
the leading and next to leading logarithms to all orders in \as.  
It has been  shown in ref.~\cite{O-colfac} that such calculations may be 
used to derive \as\ at \LepI, but that it is necessary also to include 
a sub-leading term of the form $G_{21} \as^2 L$ in order to achieve a 
good description of the data.  
\item[Combined NLLA+\oaa] 
The most complete embodiment of our present knowledge of QCD comes from
combining the \oaa\ and NLLA calculations.  It is necessary to 
match the calculations in such a way as to eliminate double counting of terms, 
and there are several ways of doing this.  These have been studied at \LepI,
based on which we choose the
``$\ln R$" matching scheme for the present work.  
\end{description}

To assess the range of validity of these calculations at \LepII\ 
we proceed in the following empirical manner.  
We have generated distributions of 
the five observables, $(1-T)$, $M_H$, $B_T$, $B_W$ and $y_{23}^{(D)}$, 
at the parton level, 
using  the \Jetset7.4 
parton shower model without initial state radiation.
We can assume that the data, after correction for detector acceptance, 
the effect of selection cuts and background, and hadronization, would
closely resemble these distributions.
For each observable, we then determine the 
largest range for which the theoretical calculations 
reproduce those from \Jetset
with an acceptable $\chi^2/{\bf DOF}$.  
The results are summarised in Table~\ref{tab-drw3}.
We note that the \oaa\ calculations may (in most cases) 
be extended to lower values of the observables by fitting \xmu.  
The NLLA or combined calculations allow a description down to still lower
values, but, particularly in the case of the pure NLLA calculations, 
the higher values of the observables are less well modelled.  
The NLLA and combined calculations for $B_W$ tend to give a rather poor
description of the \Jetset\ ``data" (as seen at \LepI).
The pure NLLA calculations are not applied to $y_{23}^{(D)}$,
since they are known to be incomplete, and in fact yield a poor 
fit to the \Jetset\ distributions.
\begin{table}[htb]
\begin{center}
\begin{tabular}{|r||c|c|c|c|}
\hline
Observable & \oaa\ (\xmu=1)  & \oaa\ (\xmu\ fitted)  & 
pure NLLA & Combined \oaa+NLLA \\ \hline
$(1-T)$       &  0.09--0.3 & 0.05--0.3 & 0.02--0.17 & 0.02--0.3      \\
$M_H  $       &  0.20--0.55 & 0.14--0.55 & 0.10--0.35 & 0.14--0.55      \\
$B_T  $       &  0.11--0.3 & 0.10--0.3 & 0.05--0.18 & 0.05--0.22      \\
$B_W  $       &  0.06--0.26 & 0.06--0.26 & 0.02--0.12 & 0.05--0.17      \\
$y_{23}^{(D)}$ & 0.015--0.2 & 0.005--0.2 & -- & 0.005--0.2      \\ \hline
\end{tabular}
\caption{Approximate ranges of applicability of various types of QCD 
calculation.}
\label{tab-drw3}
\end{center}
\end{table}

If, for example, we require that the hadronization corrections lie between
0.8 and 1.2, that the \Zqq\ acceptance be greater than 50\% and that the 
\WWqqqq\ contamination be less than 50\%, the regions where the data 
can be used reliably would be roughly  0.03--0.2 for 
$(1-T)$,  0.15--0.4 for
$M_H  $,  0.06--0.2 for       
$B_T  $,   0.03--0.18 for    
$B_W  $ and    0.005--0.09 for    
$y_{23}^{(D)}$. 
By comparison with Table~\ref{tab-drw3} it is evident that the regions 
in which reliable data may be obtained are best matched by the regions
in which the combined NLLA+\oaa\ calculations are valid.  Since these
are also the most complete calculations, this would appear to be the 
most promising approach.

\begin{figure}[hbtp]
\centerline{
\epsfig{figure=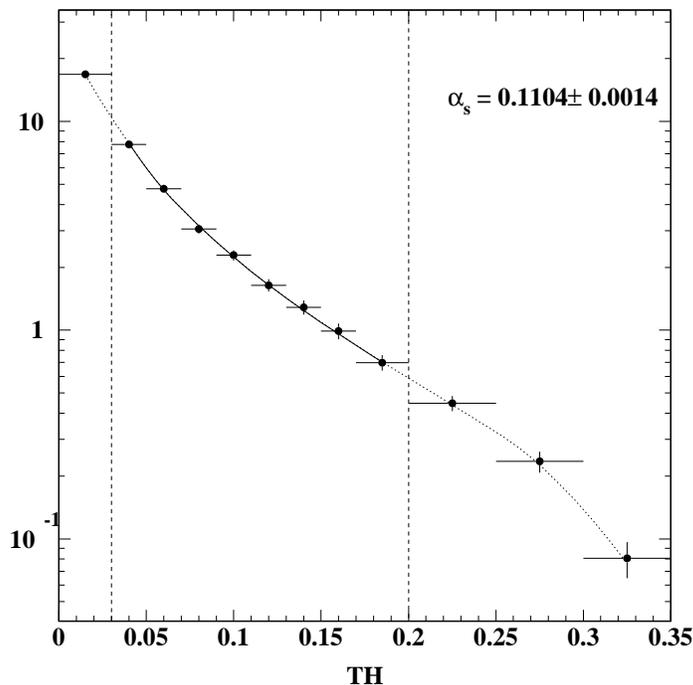,width=9cm}
%\epsfysize=8cm
%\epsffile[20 150 550 650]{drw5.ps}
}
\caption{Typical fit of the\oaa+NLLA QCD calculations to $(1-T)$
in order to determine \as.  The dotted lines delimit the fit region.}
\label{fig-drw5}
\end{figure}

We next assess the precision on \as\ which could be achieved using 
500~pb$^{-1}$ of data at \LepII.  In order to do this, we take the 
\Jetset7.4 parton level distribution, with statistical errors 
corresponding to this integrated luminosity (approximately 6500 \Zqq\ events).
We then fit the QCD theory to infer \as, fitting in the range of 
the observable given by the overlap of the ranges in 
Tables~\ref{tab-drw3}~and the regions where reliable data may be obtained. 
A typical fit (of the \oaa+NLLA calculations to $(1-T)$)
is shown in Fig.~\ref{fig-drw5}.
%%%%%%%%%%%%%%%%%%%%%%%%%%%%%%%%%%%%%%%%%%%%%%%%%%%%% 
%%%%%%%%%%%%%%%%%%%%%%%%%%%%%%%%%%%%%%%%%%%%%%%%%%%%% 
We find that the \oaa\ calculations yield typical  
statistical errors of $\pm 0.0024$, which are larger than the 
NLLA and combined \oaa+NLLA calculations (typically $\pm 0.0015$) 
because the former are only applicable towards the 3-jet region, where the
few events are found.  It also appears that the statistics are
generally insufficient to permit a precise determination of 
the scale factor $\xmu$ for the \oaa\ fits.  The pure NLLA and 
combined \oaa+NLLA calculations both appear to be competitive, 
and offer the possibility of measuring \as\ with a statistical
precision of around $\pm 0.0015$.  For the event shapes 
$(1-T)$, $M_H$, $B_T$ and $B_W$, the NLLA tend to yield smaller values
of \as, and the \oaa\ calculations larger values; the same trend was
noted at \LepI~\cite{O-colfac}.  

As at \LepI, the combined \oaa+NLLA method will probably be the preferred 
technique, because it represents the most complete theoretical
calculations, and allows the largest fraction of the data to be included 
in the analysis.  For the discussion of possible systematic uncertainties,
we therefore focus on these calculations.  In ref.~\cite{O-asnlla},
for example, 
a wide range of systematic effects were investigated.  The largest 
contribution was found to arise from variation of the renormalization scale
factor \xmu.  Other significant effects arose from varying the hadronization
model, particularly from the use of the \Herwig\ model, and from 
the influence of b-quark mass effects.  We have estimated the systematic 
errors resulting from these effects at 175~GeV, and compared with the \LepI\
experimental results.
\begin{itemize}
\item
The renormalization scale factor is varied in the range 0.5$< \xmu <$2.0.
The changes in \as\ are highly correlated with those at \LepI, though 
about 20\% smaller on average.   If we assume that it makes sense to choose the
same scale factor at \LepII\ as at \LepI, then the effective systematic 
uncertainty on the {\em change} in \as\ between \LepI\ and \LepII\ 
would be about $\pm 0.0015$ for $(1-T)$, 
$\pm 0.0025$ for $M_H$,  $\pm 0.0015$ for $B_T$, $\pm 0.0005$ for $B_W$
and $\pm 0.0003$ for $y_{23}^{(D)}$.
\item
The influence of b-quark mass effects may be crudely accounted for by
basing the parton level distributions in the correction procedure 
only on udsc quark events.  At \LepI\ this correction was found to increase
\as\ by about 0.002 for most observables.  Not surprisingly, the effect 
is much smaller at \LepII.  However, the relevant point is the difference 
between the \LepI\ and \LepII\ uncertainties, which is of the order of 0.002
(somewhat larger for $B_T$ and smaller for $M_H$). 
\item
The \Herwig\ model offers a quite different hadronization scheme from \Jetset.
Since the hadronization corrections are smaller at \LepII\ than at \LepI,
we would expect the uncertainty associated with the use of different
models to be reduced.  This is generally the 
case, but the correlation between the \Herwig\ uncertainties
at \LepI\ and \LepII\ is unclear.  This is partly because different fit regions
have been used, and also different versions of the models.  Clearly, in order
to establish a reliable systematic uncertainty on the difference in \as\
between \LepI\ and \LepII\ it would be necessary to make a more careful
analysis using consistent versions of the models at the two energies.
For some observables at least (e.g. $B_W$ and $y_{23}^{(D)}$) it seems 
plausible that the hadronization uncertainty could be quite small.
\end{itemize}

In summary, it appears that systematic errors
would not preclude making a useful measurement of the {\em difference} in
\as\ between \LepI\ and \LepII.
The renormalization scale uncertainty seems to be comparable with or smaller
than
the statistical error.  The uncertainty associated with b-quark mass effects 
could perhaps be reduced by further analysis and theoretical work.
The uncertainties associated with the choice of hadronization models are
less clear; it may be necessary to reanalyse the \LepI\ data using the
same models and parameter sets as employed in the \LepII\ analysis, 
and the same fit regions, in order to minimise the uncertainties.
Nonetheless, it seems that the systematic errors could be quite small,
for some observables at least (especially $B_W$ and $y_{23}^{(D)}$,
according to our study).  It may be noted that recent studies of 
non-perturbative (power) corrections to the mean values of event shape
observables~\cite{bib-dokweb} suggest that certain observables
or combinations of observables 
might be expected theoretically to have especially small hadronization
uncertainties (e.g. $y_{23}^{(D)}$ or $T-2C/3\pi$~\cite{bib-webber2}).

\section{Event Shapes -- Theoretical\protect\footnote{Written by
P. Nason, including contributions of M.H. Seymour, N. Glover and K. Clay.}}
\label{Nason}
Since the completion of the Yellow Report for LEP1 \cite{Yellow1}
much progress has been achieved in the theoretical calculations of shape
variable distributions.
A technique of resummation of contributions enhanced near the two--jet region
has been studied and fully implemented in refs.
\cite{bcm80,jp_CTTW,jp_CDOTW,jp_fact,jp_CT,jp_greco,jp_tesima,
jp_hjm,jp_EEC,jp_broad,jp_jmult,jp_disjm}.
Furthermore, new calculations of shape variable distributions
(implemented as computer code) have become available.

Calculation of shape variables are all based upon the original work of
ref.~\cite{ERT}. This calculation was also performed in ref.~\cite{GKSS}.
Although the analytic results did agree, several problems where found
in the comparison of numerical results (see ref.~\cite{Yellow1}
for a small review).
While at the time of ref. \cite{Yellow1} it was hard to find precision
calculations of jet shape distributions that agreed with each other,
today we have at least three general purpose programs that do agree.
One, the program EVENT, was developed for ref.~\cite{Yellow1}.
Results of shape variables distributions performed with this program
are reported there, and have served as a benchmark for comparison with other
computations. In ref.~\cite{Glover} a new computation was performed,
which agrees with good accuracy with ref.~\cite{Yellow1}. Furthermore,
very recently, yet another calculation was completed \cite{Seymour}.
In ref.~\cite{Seymour} also oriented events are implemented,
and apparently they will also be implemented in ref.~\cite{Glover}.
This means that it will be possible to compute distribution of shape variables
that do depend upon the orientation of the incoming beams axis,
unlike all shape variables that where used up to now (see the next section).

The most disturbing disagreement on shape variables was found to be on
the Energy-Energy correlation (EEC). The computation performed in
ref.~\cite{SEllis1} was found in important disagreement with other
calculations, and in particular with ref.~\cite{Yellow1}.
Recently, in ref.~\cite{SEllis2} the calculation of ref.~\cite{SEllis1}
was repeated. The result of the new calculation was found in disagreement
both with the result of ref.~\cite{Yellow1} and with ref.~\cite{SEllis1}.
No clear statement is made in ref.~\cite{SEllis2} upon the origin of the
discrepancy. It is however claimed that the disagreement comes
from the region in which besides the quark--antiquark pair, two soft
gluons have been radiated.
The EEC is in fact peculiar, in the sense that even configurations
with thrust near 1 can contribute to the EEC at angles far away from
$0$ and $\pi$.
Because of the lack of a more complete theoretical paper form the authors
of ref.~~\cite{SEllis2}, we thought that the most useful thing to be done
for the present report is to perform a high--precision comparison
of the different computations of the EEC, that can serve as benchmark
for future calculations.
In order to achieve high precision, instead of computing the EEC
itself as a function of the angle, we computed its moments.
The energy-energy correlation is defined as
\beq
\mbox{EEC}(\chi) =\frac{1}{\sigma}\sum_{ij}\;
\int d^3\vec{p}_i\, d^3\vec{p}_j\;\frac{d\sigma}{d^3\vec{p}_i\, d^3\vec{p}_j}
\frac{E_i\,E_j}{E^2}\;\delta(\vec{p}_i\cdot\vec{p}_j-\cos\chi)\;.
\eeq
We define
\beq
\int \mbox{EEC}(\chi)\;\sin^{2+m}\chi\;\cos^n\chi\;d\cos\chi
=\frac{\as}{2\pi}A_{\rm EEC}^{(m,n)}+\left(\frac{\as}{2\pi}\right)^2
B_{\rm EEC}^{(m,n)}
+{\cal O}(\as^3)
\eeq
where $\as=\as(E_{\rm cm})$. The coefficients
$B_{\rm EEC}^{(m,n)}$ have the following colour structure
\beq
B_{\rm EEC}^{(m,n)}=C_F\left(C_A\,B_{C_A}^{(m,n)}+C_F\,B_{C_F}^{(m,n)}
+T_f\,n_f\,B_{T_f}^{(m,n)} \right).
\eeq
We then asked K. Clay (C), N. Glover (G), M. Seymour (S), and the author (N),
to compute $B_{C_A}^{(m,n)}$, $B_{C_F}^{(m,n)}$,
$B_{T_F}^{(m,n)}$ with the programs of ref.~\cite{SEllis2},
\cite{Glover}, \cite{Seymour} and \cite{Yellow1}
for $m=0,\ldots,5$ and $n=0,1$. All four computations
agreed within errors for the $B_{T_F}^{(m,n)}$ term. In the other
two cases we found disagreements. The results are reported in tables
\ref{EECComp1} and \ref{EECComp2}.
%
% This is a sample Table
%
\begin{table}[hbtp]
\begin{center}
\begin{tabular}{||c|c||c|c|c|c||}
\hline\hline
m&n& N                  & G                  & S             & C     \\
\hline
0&0&$ 50.82  \pm 0.05  $&$  50.54 \pm 0.03  $&$ 50.72  \pm 0.02  $&$ 46.4  \pm 0.2$\\
1&0&$ 35.76  \pm 0.04  $&$  35.53 \pm 0.02  $&$ 35.64  \pm 0.02  $&$ 32.09 \pm 0.06$\\
2&0&$ 28.94  \pm 0.03  $&$  28.75 \pm 0.02  $&$ 28.82  \pm 0.02  $&$ 25.73 \pm 0.04$\\
3&0&$ 24.92  \pm 0.03  $&$  24.75 \pm 0.02  $&$ 24.80  \pm 0.02  $&$ 22.03 \pm 0.04$\\
4&0&$ 22.20  \pm 0.03  $&$  22.05 \pm 0.02  $&$ 22.09  \pm 0.02  $&$ 19.54 \pm 0.04$\\
5&0&$ 20.21  \pm 0.03  $&$  20.07 \pm 0.02  $&$ 20.10  \pm 0.02  $&$ 17.74 \pm 0.03$\\
0&1&$ -6.468 \pm 0.006 $&$ -6.50  \pm 0.01  $&$ -6.455 \pm 0.005 $&$ -6.0  \pm 0.15$\\
1&1&$ -2.356 \pm 0.004 $&$ -2.365 \pm 0.009 $&$ -2.344 \pm 0.003 $&$ -2.15 \pm 0.03$\\
2&1&$ -1.189 \pm 0.003 $&$ -1.194 \pm 0.008 $&$ -1.177 \pm 0.003 $&$ -1.06 \pm 0.02$\\
3&1&$ -0.714 \pm 0.003 $&$ -0.718 \pm 0.007 $&$ -0.702 \pm 0.003 $&$ -0.62 \pm 0.01$\\
4&1&$ -0.478 \pm 0.003 $&$ -0.479 \pm 0.007 $&$ -0.466 \pm 0.003 $&$ -0.41 \pm 0.01$\\
5&1&$ -0.344 \pm 0.003 $&$ -0.344 \pm 0.006 $&$ -0.331 \pm 0.003 $&$ -0.28 \pm 0.01$\\
\hline\hline
\end{tabular}
\end{center}
\caption{Comparison of different computations
of the $B_{C_A}^{(m,n)}$ coefficients.}
\label{EECComp1}
\end{table}

%
%
% This is a sample Table
%
\begin{table}[hbtp]
\begin{center}
\begin{tabular}{||c|c||c|c|c|c||}
\hline\hline
m&n& N                 & G                 & S                  & C     \\
\hline
0&0& $-13.29\pm 0.01 $ & $-13.94\pm 0.05$  & $-13.40\pm 0.05$   & $7.2   \pm 0.2 $ \\
1&0& $-5.09 \pm 0.01 $ & $-5.38 \pm 0.04$  & $ -5.14\pm 0.04$   & $9.98  \pm 0.04$ \\
2&0& $-2.98 \pm 0.01 $ & $-3.20 \pm 0.03$  & $ -3.02\pm 0.03$   & $9.55  \pm 0.03$ \\
3&0& $-2.11 \pm 0.01 $ & $-2.29 \pm 0.03$  & $ -2.14\pm 0.03$   & $8.86  \pm 0.02$ \\
4&0& $-1.65 \pm 0.01 $ & $-1.81 \pm 0.03$  & $ -1.67\pm 0.03$   & $8.22  \pm 0.02$ \\
5&0& $-1.36 \pm 0.01 $ & $-1.51 \pm 0.03$  & $ -1.39\pm 0.03$   & $7.69  \pm 0.02$ \\
0&1& $ 4.906\pm 0.002$ & $ 4.92 \pm 0.01$  & $  4.892\pm 0.006$ & $2.6   \pm 0.2 $ \\
1&1& $ 0.240\pm 0.002$ & $ 0.259\pm 0.008$ & $  0.232\pm 0.004$ & $-0.58 \pm 0.02$ \\
2&1& $-0.383\pm 0.002$ & $-0.367\pm 0.006$ & $ -0.386\pm 0.004$ & $-0.80 \pm 0.01$ \\
3&1& $-0.458\pm 0.002$ & $-0.445\pm 0.005$ & $ -0.459\pm 0.003$ & $-0.72 \pm 0.01$ \\
4&1& $-0.428\pm 0.002$ & $-0.417\pm 0.005$ & $ -0.429\pm 0.003$ & $-0.606\pm 0.007$\\
5&1& $-0.381\pm 0.002$ & $-0.371\pm 0.004$ & $ -0.381\pm 0.003$ & $-0.510\pm 0.006$\\
\hline\hline
\end{tabular}
\end{center}
\caption{Comparison of different computations of the
$B_{C_F}^{(m,n)}$ coefficients.}
\label{EECComp2}
\end{table}
It is clear that the results N, G and S agree with each other with high
accuracy, while C is seriously different. Observe that, although for all
practical purposes N, G and S agree with each other, there are among them
discrepancies of several standard deviations. We attributed these differences
as an underestimate of the errors, rather than to a real difference in the
calculation. More details on the different characteristics of the three
computer codes are given in the generator's section \cite{MHSevgen}.

\section{Next-to-leading Order Calculations of Oriented Event
Shapes\protect\footnote{Author: M.H. Seymour}}
\label{Seymour}
At LEP2, it will become increasingly important to be able to cut out
some angular regions to control the backgrounds, and to define event
shapes that are invariant under boosts along the beam direction to study
continuum events with initial-state radiation.  To make predictions for
such quantities it is essential to use the full matrix elements for
$\mathrm{e^+e^-\to q\bar{q}g},$ including the full Z$/\gamma$
interference and the polarisation of the exchanged boson.  Two programs
have recently become available that include these matrix elements,
EERAD\cite{MHSgg} and EVENT2\cite{MHScs}.  These use completely
different methods to implement the cancellation of poles between real and
virtual contributions, as described in \cite{MHSevgen}, but the results
are in excellent agreement with each other.

\begin{figure}
  \centerline{\epsfig{figure=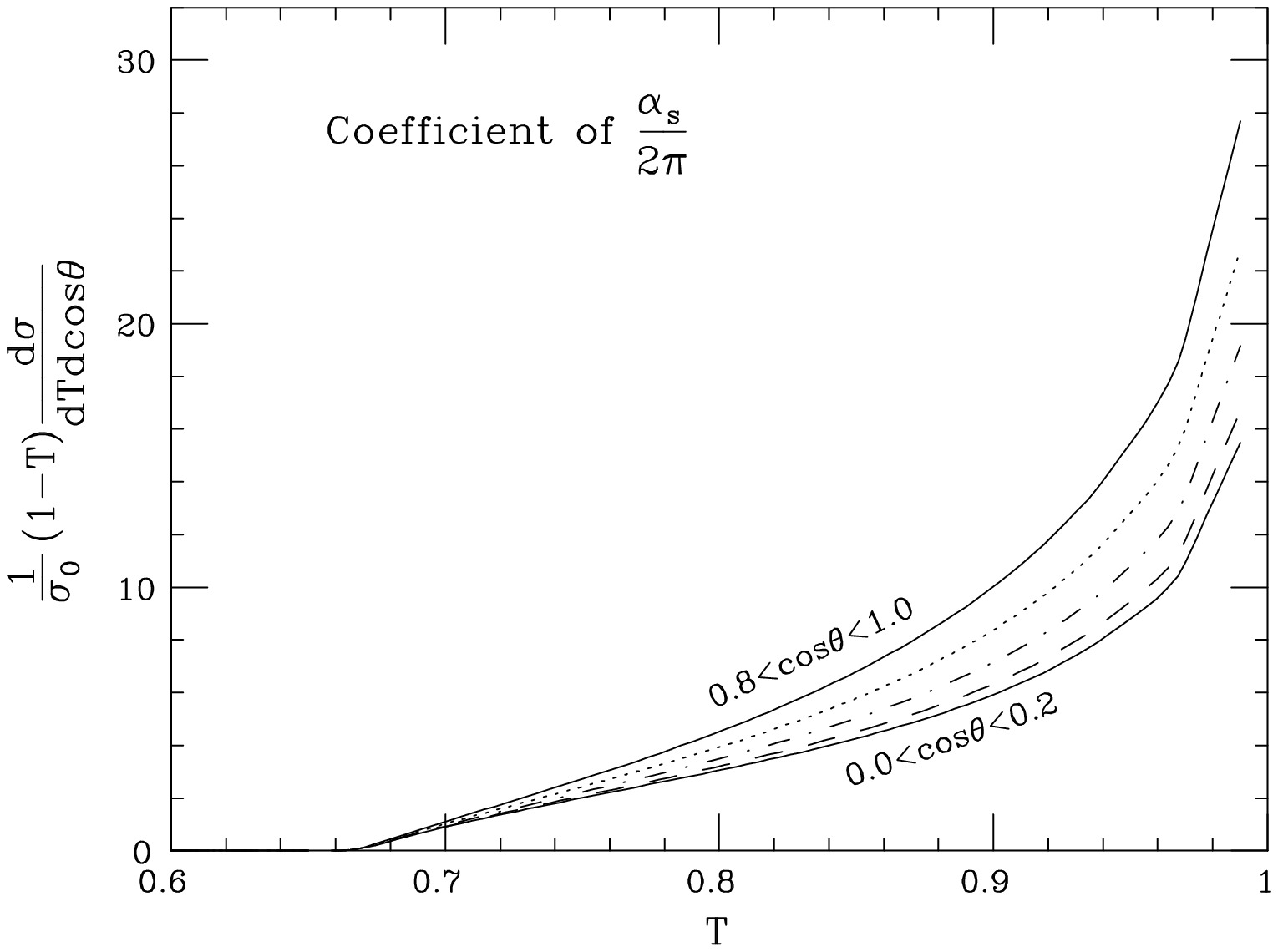,height=6cm}\hspace{\fill}
              \epsfig{figure=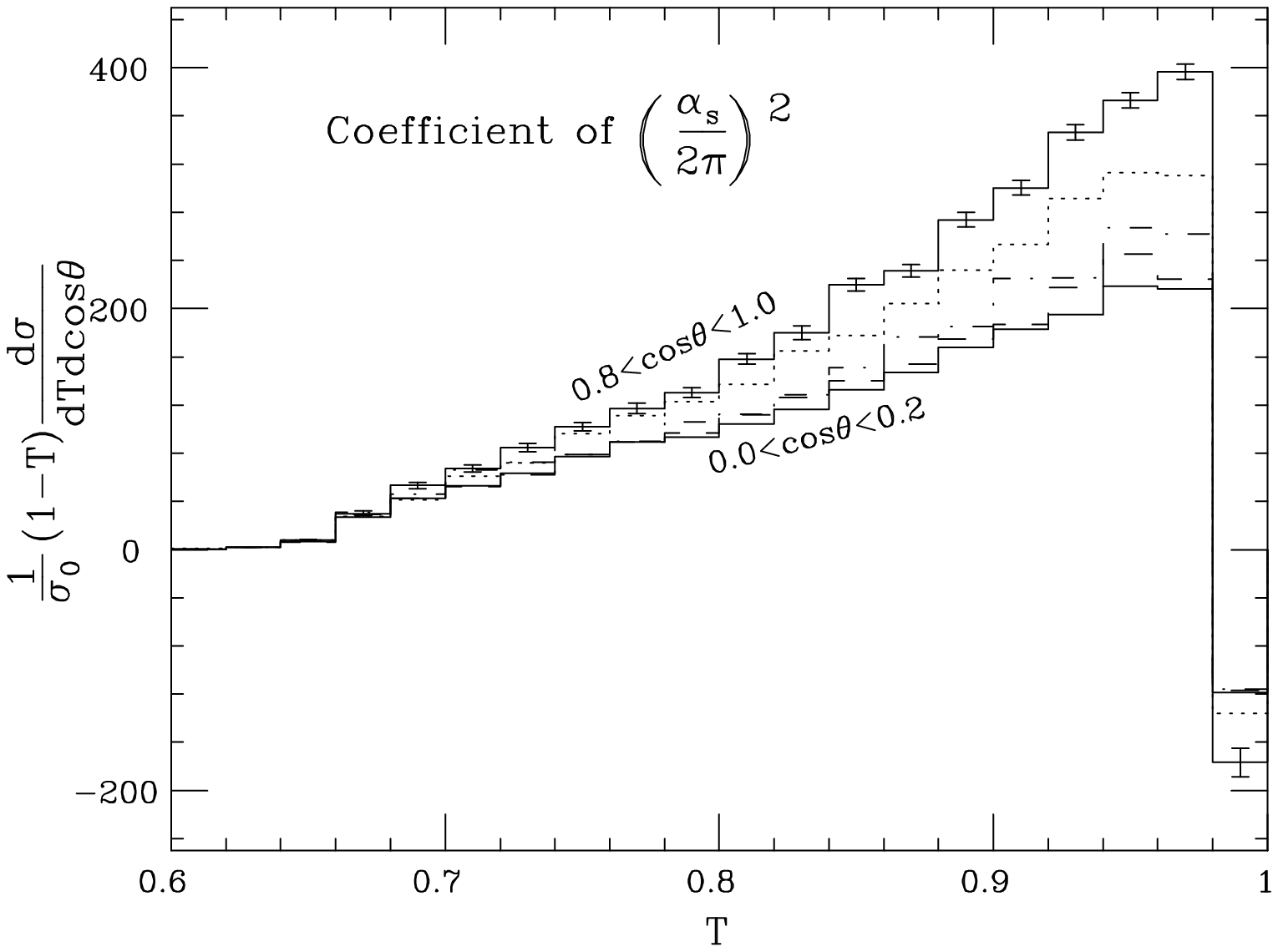,height=6cm}}
%  \vspace*{6cm}
  \caption[]{The coefficients of the thrust distribution for five
    bins in $\cos\theta,$ where $\theta$ is the angle between the thrust
    axis and the beam.  The errors shown are purely statistical and are
    similar for each histogram, so we only show them for one.}
  \label{MHSfig1}
\end{figure}

\begin{figure}
  \centerline{\epsfig{figure=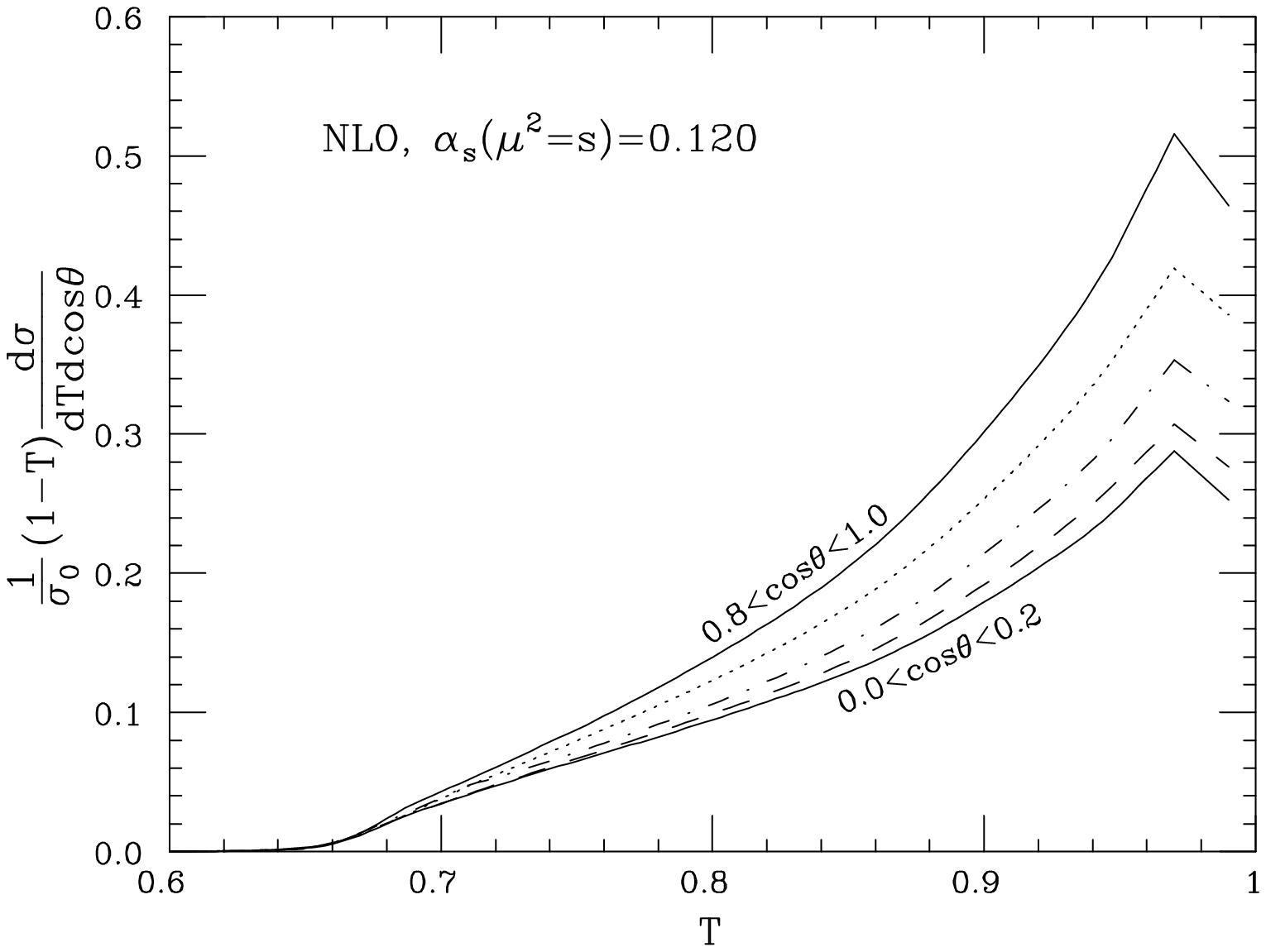,height=6cm}\hspace{\fill}
              \epsfig{figure=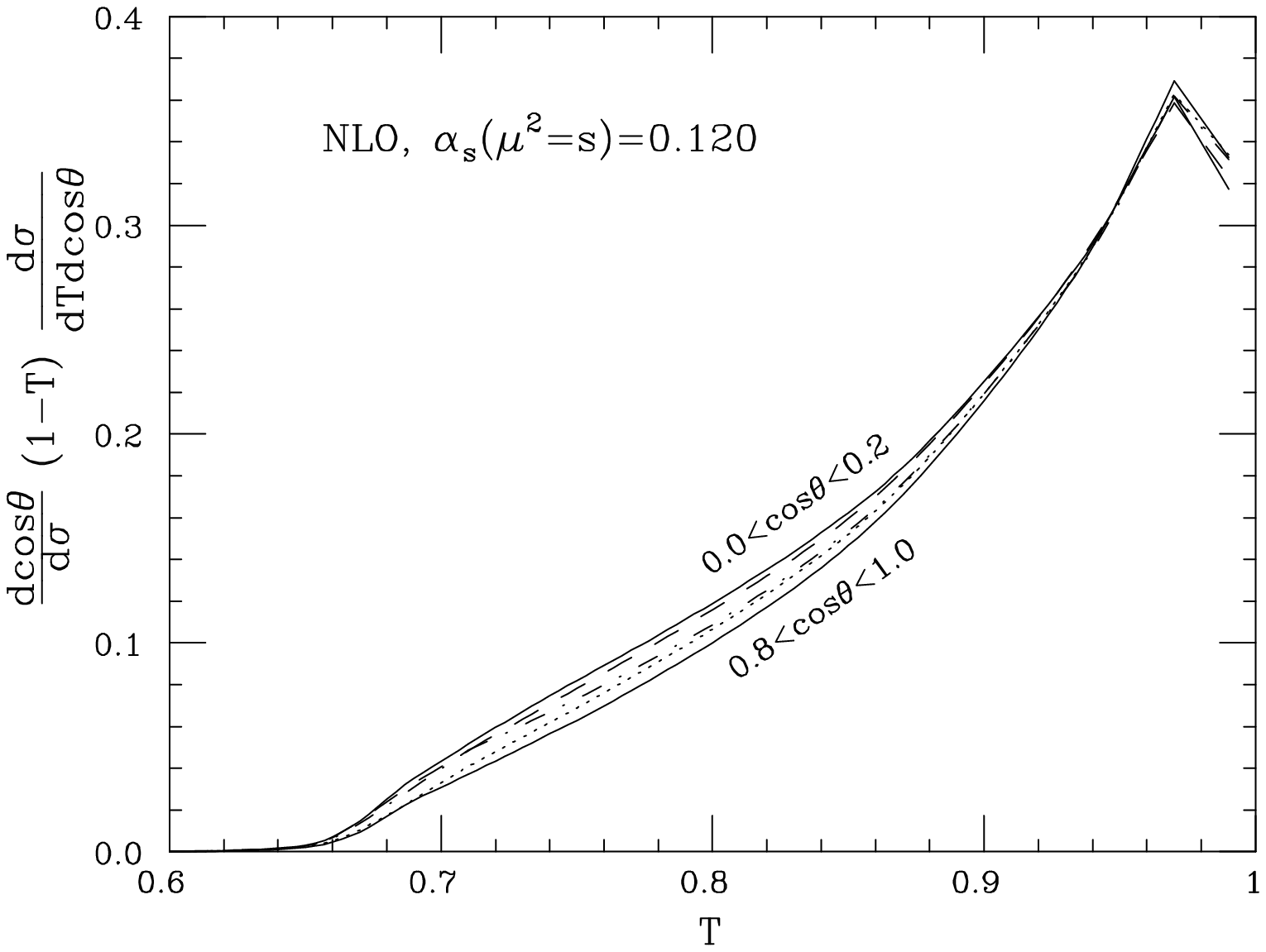,height=6cm}}
%  \vspace*{6cm}
  \caption[]{Predictions for the thrust distribution for five bins in
    $\cos\theta$ normalized to (a) the total number of events and (b)
    the number of events in each bin.}
  \label{MHSfig2}
\end{figure}

As an example of an oriented event shape we study the thrust
distribution as a function of the thrust axis direction.  As
usual\cite{MHSlep1}, we parametrize the distribution as
\begin{equation}
  \frac1{\sigma_0} (1-T) \frac{d\sigma}{dT d\!\cos\theta} =
  \frac{\alpha_s(\mu^2)}{2\pi}A(T,\cos\theta)
  +\left(\frac{\alpha_s(\mu^2)}{2\pi}\right)^2
  \left[A(T,\cos\theta)2\pi b_0\log\frac{\mu^2}{s} +
    B(T,\cos\theta)\right]\!.
\end{equation}
The definition of thrust has a forward-backward ambiguity, so we are at
liberty to define $\cos\theta>0$.  The leading order term is known
analytically\cite{MHSgary},
\begin{eqnarray}
  A(T,\cos\theta) &=& C_F\left[
  \left\{\frac{2(3T^2-3T+2)}{T}\log\frac{2T-1}{1-T} -
      3(3T-2)(2-T)\right\}\mbox{$\frac34$}(1+\cos^2\theta)
    \right.\nonumber\\&&\phantom{C_F}\;\;\left.
    + \left\{\frac{2(3T-2)(2-T)(1-T)}{T^2}\right\}
      \mbox{$\frac34$}(1-3\cos^2\theta)\right].
\end{eqnarray}
Integration over $\cos\theta$ immediately gives us the expression in
\cite{MHSlep1}.  Numerical results for $A$ and $B$ are shown in
Fig.~\ref{MHSfig1}.
Combining these coefficients with an $\alpha_s$ value and factorisation
scale choice, $\alpha_s(\mu^2=s)=0.120,$ we obtain the predictions shown
in Fig.~\ref{MHSfig2}a.
Alternatively, we can divide out the trivial dependence on $\cos\theta$
by normalising each curve to the number of events in that bin, given
by\cite{MHSgary}
\begin{equation}
  \frac1{\sigma_0} \frac{d\sigma}{d\!\cos\theta} =
  \mbox{$\frac34$}(1+\cos^2\theta)\left\{1+\frac{\alpha_s}{\pi}\right\}
  +\frac{\alpha_s}{\pi}
    \left\{8\log\frac32-3\right\}(1-3\cos^2\theta)
\approx
  \mbox{$\frac34$}(1+\cos^2\theta)
  +\frac{\alpha_s}{\pi},
\end{equation}
where the approximation is good to better than 1\%.  The result is shown
in Fig.~\ref{MHSfig2}b, where we see that the majority of the
$\cos\theta$ dependence in Fig.~\ref{MHSfig2}a was from this dependence
of the total event rate and the residual dependence is rather small.
Nevertheless, it should be measurable with the full statistics of LEP1.

\section{Fragmentation functions\protect\footnote{Author: C. Padilla.}}
\label{Padilla}
The measurement of fragmentation functions at different energies
and the comparison with the theoretical predictions, either implemented 
in the Monte Carlo programs or deduced from other measured data,
can be used to perform different QCD tests and to tune the parameters 
describing the fragmentation processes inside the Monte Carlo programs. 

At the energies available at LEP II the scaled energy 
($x\equiv 2E/\sqrt{s}$) distributions 
for charged particles can be measured for 
${\mathrm q}\overline{\mathrm q}$\/ events in which the mass
of the hadronic system
is close to the 
centre--of--mass energy of the collision.
Furthermore, the fragmentation function of the W boson can also be measured
and compared to the expectation that comes from the measurement of the
fragmentation functions for different enriched flavour samples at LEP I, after
correcting for the small scaling produced for the different masses of the
Z and the W boson and for the different flavour composition.

This Section describes how the measurement of the 
scaled energy distributions 
can be made and what can be expected in the 
measurement of $\alpha_s$\/ from scaling violations.

\subsection{Measurement of scaled energy distributions}

The measurement of the charged scaled energy distributions will follow
the same procedure used at lower centre--of--mass energies. 
At centre--of--mass energies of the Z mass, hadronic events can be selected
with very high purity and small backgrounds (coming mainly from $\tau$ events).
At LEP II centre--of--mass energies, most of the $q\bar{q}$
events (more than 75\%) will 
radiate an initial state hard photon such a way that the effective 
centre--of--mass energy of the collision will be reduced to below 120~GeV.
These events have a high boost along the collision axis and have to be removed.

The selection of hadronic events will follow a procedure very similar 
to the one presented in subsection~\ref{sect-Zsel}. After some minimal 
requirements on track quality, number of tracks and total 
measured energy of these 
tracks, additional selection variables have to be considered. 
Good containment of the events can be obtained with cuts in 
the sphericity or thrust axis. 

Monte Carlo simulations performed in ALEPH, based upon DYMU3 and
JETSET, including full simulation of the detector
response, show that requiring a visible
mass of the event above 120~GeV and a normalized 
balanced momentum of the charged tracks along the beam axis below 0.3, 
a selection efficiency of $\sim 18\%$\/ can be achieved. 
The percentage of selected events such that the invariant mass 
of the propagator is below 120~GeV is reduced to approximately 7\% with 
this selection procedure. 

The backgrounds from dilepton events are small at this 
level. However, the background from WW events could still be substantial.
A cut in missing momentum will remove most of the events
in which one of the W has decayed leptonically. The remaining events
in which both W decay hadronically can be removed by considering
appropriate shape variables. Events resulting from the fragmentation
of two W bosons will have a four-jet topology that makes them more
spherical than the ones resulting from 
$\mathrm Z/\gamma \rightarrow {\mathrm q}\overline{\mathrm q}$.
In subsection~\ref{sect-Zsel} a discussion of the various possible approaches
is given.
For the case of the measurement of the scaled energy distributions, 
a cut on thrust $T>0.925$ would be appropriate, since
(unlike the case of shape variables) such a cut does not introduce
strong biases in the shape of the fragmentation function.

The whole selection procedure should result in a cross section for
${\mathrm q}\overline{\mathrm q}$\/ events of 
$~\sim 11$~pb with less than 1\% of events with the 
effective centre--of--mass energy 
below 120~GeV and with a background of WW events below 5\%. Assuming an
integrated luminosity of 500~pb$^{-1}$\/ 
the expectation is to have $\sim 6000$\/ selected 
hadronic events. 

It can be assumed that the background can be subtracted
statistically using Monte Carlo techniques, and that 
the distribution is corrected using a hadronic event generator 
(with parameters adjusted to describe the data) for the 
effects of geometrical acceptance, detector efficiency and resolution, 
decays of long-lived particles (with $\tau>1$~ns), 
secondary interactions and residual initial state photon radiation. 
The bin-to-bin correction factors are below 10\% using the selection
described above. 

\begin{figure}[hbtp]
\vspace{0.1cm}
\centerline{
\epsfig
{figure=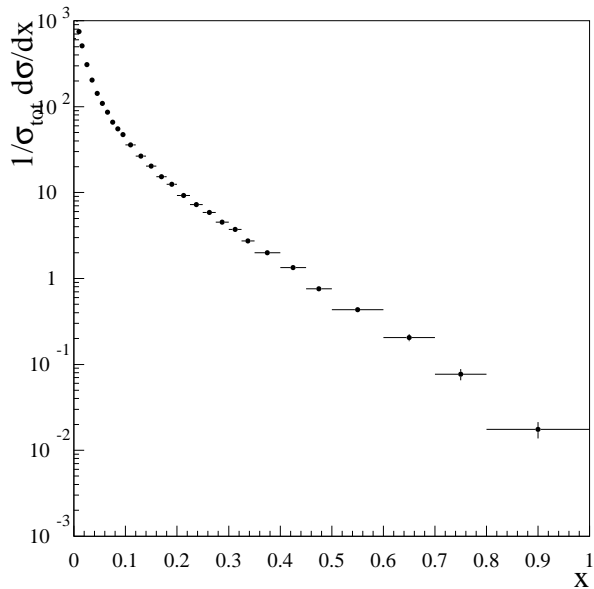,height=8cm,bbllx=0cm,bblly=5.5cm,bburx=21cm,bbury=23cm,angle=0}
}
\caption{\it Scaled energy distribution for hadronic events at a 
centre--of--mass energy of 180~GeV according to the JETSET Monte Carlo. 
The error bars correspond to the statistics of 6000 events.}
\label{scal_en_dist_qq}
\end{figure}

Figure~\ref{scal_en_dist_qq} shows the Monte Carlo scaled energy 
distribution for the statistics of 6000 events. The energies of the 
particles before detector effects have been used
to construct the distribution. Additional systematic uncertainties coming
from possible 
discrepancies between the real detector performance and the simulated one
and from the  dependence on the hadron production model used to correct 
the data  for detector effects will have to be considered.

The measurement of the W fragmentation function will require the selection
of hadronic W events. 
The events in which one of the W decays leptonically can be selected
using missing momentum or tagging a high-momentum lepton. The rest 
of the particles can be used to determine the momentum  
of the hadronically decaying W boson and to construct the
scaled energy distribution, after boosting the particles into the 
rest frame of the parent W boson. 
In the case that both W bosons decay hadronically the techniques 
used in the measurement of the W mass can be used to unambiguously assign
the jets to the corresponding W bosons.

\subsection{Scaling violations: QCD tests}

The analysis of scaling violations with the data available at LEP and 
data from lower centre--of--mass energy experiments 
(PEP, PETRA, TRISTAN) has focused on
the measurement of $\alpha_s$~\cite{aleph_scal,delphi_scal}.
The prediction of scaling violations in fragmentation functions
of quarks and gluons is similar to that predicted in 
structure functions in 
deep-inelastic lepton-nucleon scattering.

In an electron-positron collider, scaling violations 
are observed  in the dependence of
the distribution of the scaled energy 
of final-state particles in hadronic events on the centre--of--mass 
energy $\sqrt{s}$. This comes about because with 
increasing $\sqrt{s}$\/ more phase space for gluon radiation and thus 
for final-state particle production becomes available, leading to a softer
$x$-distribution. As the probability for gluon radiation is proportional to 
the strong coupling constant, a measurement of the scaled-energy 
distributions at different centre--of--mass energies compared to the QCD 
prediction allows one to determine the only free parameter of 
QCD, $\alpha_s$.
A recent review of the relevant theoretical ideas has been given in
ref.~\cite{ref:NASON}. For another recent theoretical analysis %of the data
see~\cite{Skachkov}. 

A reliable measurement of scaling violations has 
to disentangle the true QCD evolution
from effects due to the dependence of the flavour composition
upon the centre--of--mass energy.
Since heavy flavours, after their decay into light particles,
typically have softer fragmentation functions,
when going from centre--of--mass energies below the Z mass
towards the Z mass, the $b$ content increases, and it decreases
again when going towards higher energies.
To analyse the data in a model independent way, final-state 
flavour identification and a measurement of the gluon fragmentation function 
are needed.
This procedure has been followed in the analysis performed in 
ref.~\cite{aleph_scal}, where enriched $uds$-, $c$-, and $b$-quark
scaled energy distributions, together with the measurement of the gluon
fragmentation functions and the longitudinal cross section, have been used
to constraint the fragmentation functions for the different flavours and the
gluon. It was assumed that the fragmentation functions of 
$u$, $d$, and $s$\/ quarks are the same.
In the analysis presented there, a total of 15 parameters besides
$\alpha_s$\/ are fitted to all the available. The parameters contain 
information on the fragmentation functions for the different quarks and
the gluon and also a parametrisation of the 
non-perturbative contributions to the evolution.
The value of the strong coupling constant obtained from this fit is
\begin{equation}
   \asmz = 0.126 \pm 0.007(exp) \pm 0.006(theory) = 0.126 \pm 0.009 \mm{5}.
\end{equation}
The experimental error is the result of the combination in quadrature 
of the errors  from the fit (0.0053), the uncertainties in the 
flavour composition of the enriched scaled energy distributions and 
the assumptions on the normalisation errors for those low-energy experiments
where this error is not specified. 
The theoretical error is estimated by varying 
the factorisation and renormalisation scales.

\begin{figure}[hbtp]
\vspace{0.1cm}
\centerline{
\epsfig
{figure=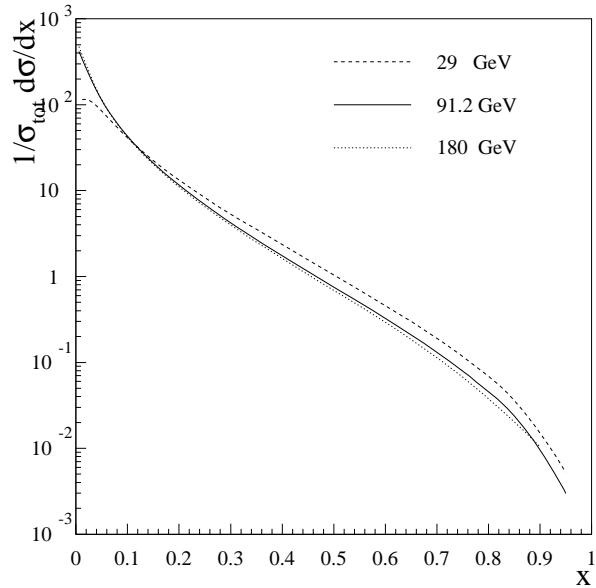,height=8cm,bbllx=0cm,bblly=5.5cm,bburx=21cm,bbury=23cm}
}
\caption{\it Result of the scaling violation fit to the distributions at 
centre--of--mass energies at 29~GeV, 91.2~GeV and 180~GeV. 
}
\label{fit_result}
\end{figure}

A possible extension of this analysis has been investigated by including the 
predicted distribution measured at a centre--of--mass energy of 180~GeV
(figure~\ref{scal_en_dist_qq}). Figure~\ref{fit_result} shows the result
of the fit to the scaled energy distributions at
three centre--of--mass energies
(29~GeV, 91.2~GeV and 180~GeV). The fact that the variations with energy
of the fragmentation functions is logarithmic makes the difference between
the distributions at 180~GeV and 91.2~GeV smaller than that between 
91.2~GeV and 29~GeV. This is accentuated by the fact that the flavour
composition changes between 91.2~GeV and
180~GeV, in particular the percentage of $b$\/ quarks
diminishes when going to energies above the Z pole. Since the fragmentation
function for $b$\/ quarks is softer, this hardens the inclusive 
distribution at LEP II energies.

The error in $\asmz$\/ coming from the fit is not improved by
including the distribution measured at 180~GeV. It was found, however, that
with four times the predicted available statistics, a 10\% improvement in this
error could be obtained. The conclusion is that the analysis could serve 
as another consistency check of the predicted QCD scaling violations. 
Improvement in the error on $\asmz$\/ may come from several sources.
A better understanding of 
the flavour tagging algorithms used to measure the flavour-enriched
distributions could improve the experimental systematic error.
Progress on the theoretical side, for example
the extension of the formalism to describe better
the low--$x$ region (see Section~\ref{Hautman})
could also be helpful.

\begin{figure}[hbtp]
\vspace{0.1cm}
\centerline{
\epsfig
{figure=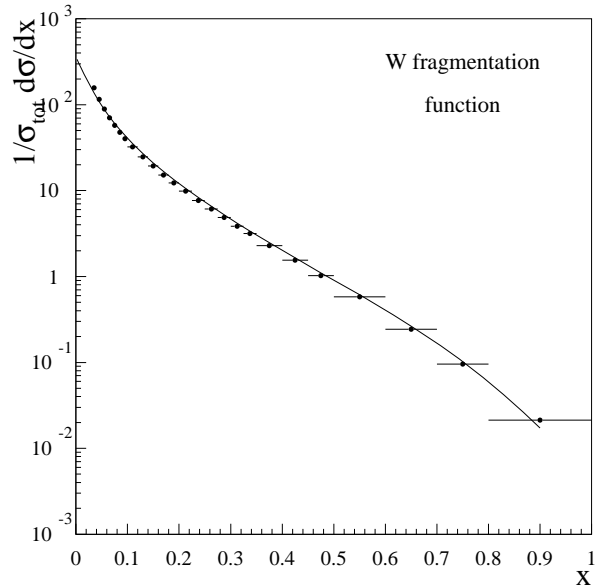,height=8cm,bbllx=0cm,bblly=5.5cm,bburx=21cm,bbury=23cm}
}
\caption{\it W fragmentation function predicted by the PYTHIA Monte Carlo
(points), compared with the QCD prediction resulting from the analysis of
scaling violations.
}
\label{w_frag}
\end{figure}

Another consistency check can be performed by using the measured
flavour-enriched
distributions at the Z peak and the scaling violation formalism to 
predict the fragmentation function in W decays.
The fragmentation functions obtained from
the fit to all data for the different quark flavours can be evolved to
the mass of the W. Then the W scaled energy distribution can be
predicted using the W decay branching ratios for each flavour.
Figure~\ref{w_frag} shows, in the continuous line, the prediction
that results from this procedure. The points are the  
W scaled energy distribution as 
predicted by the PYTHIA Monte Carlo. 

\subsection{Small-$x$ fragmentation\protect\footnote{Author: F. Hautmann}}
\label{Hautman}
In the region of small values of the momentum fraction $x$, 
the behaviour of 
the fragmentation functions 
may be significantly affected by 
phenomena related to the 
coherence of soft gluon radiation (for a review of this 
subject see, for instance, Ref.~\cite{bas}).  
These 
effects 
 are expected to result in a suppression of hadron production in 
the small-$x$ (or soft) region, and to modify both the $x$-shape and the $Q^2$-dependence
of the 
inclusive single-particle spectrum.  
In particular, 
as a consequence of coherence, 
 when the  momentum fraction becomes small  
the gluon fragmentation function is expected to 
peak at a value dependent on the hard scale of the process, and 
be damped  
in the soft region.    

From the standpoint of perturbation theory,  
coherence effects show up as logarithmic corrections 
$\alpha_S^k \log^{m}(1/x)$ ($m \leq 2 k - 2$) to the splitting 
and coefficient 
functions 
which control the perturbative 
evaluation 
of the 
fragmentation functions. For example, 
the gluon splitting function $P_{g g} (\alpha_S, x) $ 
has the small-$x$ behaviour 
(${\bar \alpha}_S \equiv \alpha_S N_c / \pi$) 
\begin{equation}
\label{pggx}
P_{g g} (\alpha_S, x) \simeq 
{{\bar \alpha}_S \over x} - 
{{\bar \alpha}_S^2 \over x} \log^2 x + 
{{\bar \alpha}_S^3 \over {3 x}} \log^4 x 
+\dots \;\;\;\;, \;\; x \ll 1 \;\;.
\end{equation}
Small-$x$ logarithms are present to all orders in $\alpha_S$, and a 
systematic way to take coherence effects into account is to 
resum these logarithms to the leading accuracy, 
next-to-leading accuracy, and so on. 

The leading-log results were determined in Refs.~\cite{bcm80,lead}, and 
can be 
best
given 
in 
the moment space 
defined 
via  
the Mellin-Fourier  
transform
\begin{equation}
\label{mel}
\gamma_{gg} (\alpha_S, \omega) \equiv \int_0^1 d x  \; x^\omega \;
P_{g g} (\alpha_S, x)  
\;\;\;\;,  
\end{equation}
and the  analogous transform for any other function of $x$. 
In the moment space 
logarithmic terms appear as multiple poles at 
$\omega \to 0$, and the 
summation of the leading contributions 
${\cal O} \left( \alpha_S^k / \omega^{2 k - 1} \right)$ is 
encompassed by the formula \cite{bcmm}
\begin{equation}
\label{gamlead}
\gamma_{gg} (\alpha_S, \omega) = {1 \over 4} \left( \sqrt{\omega^2+ 
8 \, {\bar \alpha}_S} - \omega \right)  
\;\;. 
\end{equation}  
The perturbative behaviour of this formula can be obtained by 
expanding it in the coupling $\alpha_S$. The first terms of 
the expansion read as follows
\begin{equation}
\label{gampert}
\gamma_{g g} (\alpha_S, \omega) \simeq 
{{\bar \alpha}_S \over \omega} - 2 
{{\bar \alpha}_S^2 \over {\omega^3}}  + 8 
{{\bar \alpha}_S^3 \over {\omega^5}}  
+\dots \;\;\;\; \;\;, 
\end{equation}
where in the ${\cal O} (\alpha_S)$ and 
 ${\cal O} (\alpha_S^2)$ terms one may recognize  the dominant 
part at small $x$ of the standard one-loop and two-loop 
evolution kernels for the fragmentation functions (see \cite{nw}
and references therein), whilst
higher-order terms represent corrections due to coherent emission 
of soft gluons. An important feature which can be observed  
in Eq.~(\ref{gampert}) is the alternating sign of the expansion.  
As a matter of fact, this feature extends  to the whole series, and 
the  net effect of resumming all the leading logarithms  
turns out to be a damping of the  fragmentation function 
in the soft region with respect to the lowest-order prediction. 

The asymptotic properties of the resummed expression  (\ref{gamlead}) 
are conversely determined by its behaviour near $\omega = 0$. This 
 is given by  
\begin{equation}
\label{gamzero}
\gamma_{g g} (\alpha_S, \omega) \sim 
\sqrt{{{\bar \alpha}_S \over 2}}  \;\;\;\;, \;\; \omega \to 0 
\;\;\,.
\end{equation}
 Note that the all-order summation of the perturbative poles  
$  \alpha_S^k / \omega^{2 k - 1} $ 
gives rise to a finite result at $\omega = 0$, 
and introduces on the other hand
the non-analytic behaviour in $\alpha_S$
of the square-root type.

The summation of the  next-to-leading contributions 
${\cal O} \left( \alpha_S^k / \omega^{2 k - 2} \right)$ has 
also been performed \cite{nlead}.
The explicit expression of the next-to-leading 
correction 
to Eq.~(\ref{gamlead}) reads 
\begin{eqnarray}
\label{gamnlead}
\gamma_{g g}^{NL} (\alpha_S, \omega) &=& 
\gamma_{g g}^{L} + {\bar \alpha}_S \left[ -{{11} \over {12}} - 
{{N_f} \over {6 \, C_A}} +   
\left( {11 \over 4}+
{{N_f} \over {3 \, C_A}}-{2 \over 3} {{N_f \, C_F} \over {C_A^2}} 
\right) \, {{\gamma_{g g}^{L}} \over {4 \, \gamma_{g g }^{L} + \omega}} 
\right. 
\nonumber\\
&-& \left. {11 \over 12} \,  
{{\omega \, \gamma_{g g}^{L}} \over {(4 \, \gamma_{g g }^{L} + \omega)^2}} 
- {2 \over 3} \, {{N_f} \over {C_A}} \, 
{{{ \gamma_{g g}^{L}}^2} \over {(4 \, \gamma_{g g }^{L} + \omega)^2}} 
\right] \;\;\;, 
\end{eqnarray}
 where $\gamma_{g g}^{L}$ denotes the leading term (\ref{gamlead}),  
and $N_f$ is the number of flavours.
 Next-to-leading contributions do not 
alter the qualitative behaviour determined by the leading-order 
analysis, but  provide a ${\cal O} (\sqrt{\alpha_S})$ correction 
to the position of the peak in the gluon fragmentation function. 

Phenomenological studies of the soft region of the 
single-particle spectrum have been carried out in 
Ref.~\cite{yurhyper}, on the basis of modified evolution 
equations which hold in the small-$x$ regime.
The central region of the spectrum, on the other hand, is known to be 
well described by second-order perturbation theory. 
It is therefore important   
to develop a procedure in which  
 resummed contributions are consistently 
matched on to second-order perturbation theory,  
in order to get  a uniform description of fragmentation  
over the whole phase space. 

%%%%%%%%%%%%%%%%%%%%%%%%%%%%%%%%%%%%%%%%%%%%%%%%%%%%%%%%%%%%%%%%%%%%%
%% HERE IS A SKELETON AROUND WHICH YOU CAN BUILD YOUR CONTRIBUTION %%
%%%%%%%%%%%%%%%%%%%%%%%%%%%%%%%%%%%%%%%%%%%%%%%%%%%%%%%%%%%%%%%%%%%%%
%\documentstyle[12pt,epsfig]{article}
%
% This is the established standard for the page size
% which you are not allowed to reset in your macros!
%\textheight=22.0cm
%\textwidth=17.0cm
%
% This repositions the text inside the page, and the
% result is machine-dependent, so you can reset it!
%
%
% Here you can put your personal macros at will :
%\newcommand{\WW}{\mbox{$\mathrm{W^+W^-}$}}
%\newcommand{\ZZ}{\mbox{$\mathrm{ZZ}$}}
%\newcommand{\qq}{\mbox{$\mathrm{q\overline{q}}$}}
%\newcommand{\Zzero}{\mbox{${\mathrm{Z}^0}$}}
%\newcommand{\Zqq}{\mbox{$ \Zzero / \gamma \rightarrow \qq $}}
%\newcommand{\Opal}{\mbox{O{\sc pal}}}
%\newcommand{\Aleph}{\mbox{A{\sc leph}}}
%\newcommand{\Delphi}{\mbox{D{\sc elphi}}}
%\newcommand{\LepII}{\mbox{L{\sc ep}II}}
%\newcommand{\LepI}{\mbox{L{\sc ep}I}}
%\newcommand{\Pythia}{\mbox{P{\sc ythia}}}
%
%\def \as{\relax\ifmmode\alpha_s\else{$\alpha_s${ }}\fi}
%\def\cO#1{{\cal{O}}\left(#1\right)}
%\def\eV{{\rm e\kern-0.12em V}}  \def\TeV{{\rm T}\eV}
%\def\lrang#1{\left\langle #1 \right\rangle}
%\def\beq{\begin{equation}}   \def\eeq{\end{equation}}
%\def\zp#1#2#3{{\em Zeit.Phys.}~{\bf{C#1}} (19#3) #2}
%\def\pl#1#2#3{{\em Phys.Lett.}~{\bf{B#1}} (19#3) #2}
%\def\prl#1#2#3{{\em Phys.Rev.Lett.}~{\bf{#1}} (19#3) #2}
%\def\sjnp#1#2#3{{\em Sov.J.Nucl.Phys.}~{\bf{#1}} (19#3) #2}
%
% Your contribution should appear as a Section:
%
%\section{QCD}
%
%\clearpage
%
\section{Charged Particle Multiplicities\protect\footnote{Contributors: F. Fabbri and B. Poli (exp.),
Yu.L. Dokshitzer and V.A. Khoze (th.)}}
\label{Fabbri}
The study of hadron multiplicity distributions in high energy
collisions is an important topic in multiparticle dynamics
and is generally undertaken as soon as a new energy domain
becomes accessible.
It  has been always considered a valuable tool to test our 
understanding of phenomenological approaches to multiparticle 
production and, in the framework of perturbative QCD (MLLA) 
with assumption of Local Parton Hadron Duality (LPHD) \cite{LPHD},
the average charged 
multiplicity $<n_{ch}>$ and the second binomial moment of the 
distribution, $R_2 = \frac{< n(n-1)>}{< n >^2}$, are predicted to 
evolve with energy \cite{qcd}.
The measurement of the average charged multiplicity in heavy flavoured
events is also of interest to perturbative QCD and a theoretical 
discussion on this particular topic is presented in this Section.

\subsection{Accompanying Multiplicity in Light and Heavy Quark Initiated
Events}

Perturbative QCD approach predicts a suppression of soft gluon radiation
off an energetic massive quark $Q$
inside the forward cone of aperture $\Theta_0=M_Q/E_Q$
(Dead Cone)\cite{Dead}.
This phenomenon is responsible for the ``leading heavy particle effect''
and, at the same time, induces essential differences in the structure
of the accompanying radiation in light and heavy quark initiated jets.
According to the LPHD concept\cite{Book}, this should lead to corresponding
differences in ``companion'' multiplicity and energy spectra of light hadrons.

In particular, a solid QCD prediction is that the difference of companion
mean multiplicities of hadrons, $\Delta N_{Q\ell}$,
from equal energy (hardness) heavy and
light quark jets should be $W$-independent\cite{DFK,Durham} ($W$
is the energy available for soft particle production),
up to power correction terms $\propto M_Q^2/W_Q^2$.
This constant is different for $c$ and $b$ quarks and depends on the type
of light hadron under study (e.g., all charged, $\pi^0$, etc).
This is in a marked contrast with the prediction of the so called
Naive Model based on the idea of reduction of the energy
scale\cite{Naive}, $N_{Q\bar{Q}}(W)=N_{q\bar{q}}((1-\lrang{x_Q})W)$,
so that the difference of $q$- and $Q$-induced multiplicities grows
with $W$ proportional to $N(W)$.

The data\cite{data} for charged multiplicities in $b$- and $c$-quark events
are in agreement with the energy independence of $\Delta N_{Q\ell}$.
As far as the the value of multiplicity differences is concerned,
an expression for $\Delta N_{Q\ell}$ has been derived within
the MLLA accuracy\cite{Durham} assuming $M_Q\gg\Lambda$:
\beq
\label{mllaeq}
\Delta N_{Q\ell}= N_{Q\bar{Q}}(W)- N_{q\bar{q}}(W) =
- N_{q\bar{q}} (\sqrt{e}M_Q) \left[\, 1+\cO{\as(M^2)}\,\right].
\eeq
One usually consider the directly measurable quantity
\beq\label{deltaql}
\delta_{Q\ell}=\Delta N_{Q\ell}+N_D
\eeq
where $N_D$ is the average multiplicity due to the heavy quark decay.
Quantitative QCD expectation for the difference of measured
charged multiplicities $\delta_{Q\ell}$
that includes decay products of heavy hadrons, based on (\ref{mllaeq})
was obtained in \cite{MLLAhq}.
For $b$ quarks, which are only relevant for \LepII,
the MLLA estimate  $\delta_{b\ell}=5.5\pm0.8$
exceeds the experimental value $2.90\pm0.30$.

Recently an attempt has been made\cite{PK2} to improve
eq.(\ref{mllaeq}), the result of which modification
agreed with the data
``{\em significantly better than the original MLLA prediction}\/''
(W.Metzger, \cite{data}).
However, the very picture of accompanying multiplicity as induced
by a single cascading gluon, implemented in \cite{PK2},
is not applicable at the level of subleading $\cO{\as}$ effects
(see, e.g. \cite{Book}).
Therefore a reliable theoretical improvement of the QCD prediction for the
absolute value of $N_{Q\ell}$ remains to be achieved.

DELPHI has recently measured\cite{DELPHI} the number
of $\pi^0$ in $b\bar{b}$ events to be close to that in all $Z^0$ events.
The same difference should be there at \LepII.
    
\subsection{Experimental}

The main limitations at \LepII\ in this kind of study will come from the 
limited statistics and the relatively high contamination of events 
from other physical processes, absent or totally negligible at LEP I.
Hadronic decays of \WW\ pairs and highly radiative \Zqq\ events 
are expected to constitute the dominant background. 
It was shown in subsection~\ref{sect-Zsel} that this background can be reduced 
to a tolerable level, but the selection cuts needed, due to the 
particular nature of background events, will inevitably introduce 
a bias at both low and high multiplicities. 

The present study is based on the analysis of events generated with 
\Pythia\ version 5.715 at three different energies ($\sqrt{s} = 161, 
175, 192$~GeV), with hadronization parameters tuned
to \LepI\ data~\cite{O-jttune}.
The events were fully processed through the \Opal\ detector simulation
and reconstruction program chain, but the conclusions drawn here are 
believed to be practically the same for the other experiments.
The statistics used was large compared to the most optimistic assumption
on the integrated luminosity achievable at \LepII.
Following the usual convention\cite{conv}, the charged multiplicity is 
defined as the total number of all promptly produced stable charged 
particles and those produced in the decays of particles with lifetimes 
shorter than $3 \cdot 10^{-10}$ sec.
Non-radiative \Zqq\ events were selected following the criteria 
suggested in subsection~\ref{sect-Zsel}, in particular we used a combined "stage I" 
and $B_N < 0.06$ cut.
A further background reduction was obtained by rejecting events with
a Thrust value $T < 0.8$.
The selection efficiency achieved for non-radiative events,
defined as those with an $E_{isr}<1$~GeV (see subsection~\ref{sect-Zsel}), 
was higher than $82\%$ for all the considered energies.
Due to detector acceptance and quality cuts, about $9\%$ of 
the charged particles, on average, were lost in events surviving 
cuts while the predicted unbiased 
average charged multiplicity ($<n_{ch}>= 27.3$ at $\sqrt{s}=175$~GeV)
is about $14\%$ higher than the observed one. 
Both those fractions were found to be practically energy independent.
Approximately $33\%$ of the events surviving cuts are radiative,
namely with an $E_{isr}>1$~GeV, but most of them have $E_{isr}<20$~GeV.
Background from \WW\ events never exceeds the $3\%$ level.
Residual background from other sources, like \ZZ\ pairs, single W 
and single Z production, tau pairs and two-photon events was found
to be negligible.

\begin{figure}[htbp]
\vspace{0.1cm}
\centerline{
\epsfig{figure=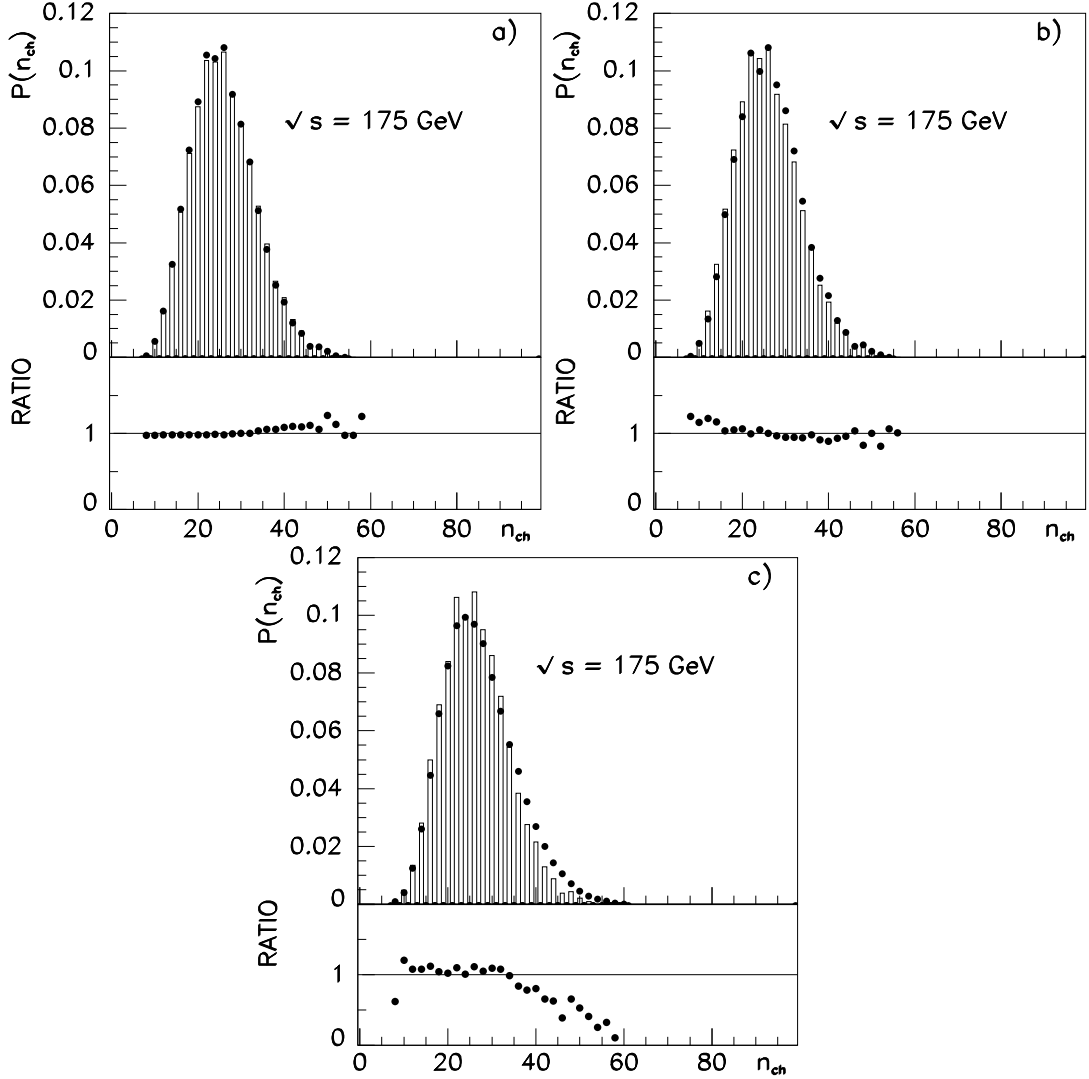,height=15cm,angle=0}
}
\caption{\it Bias from: a) residual \WW; b) radiative events; 
c) selection cuts.}
\label{mult1}
\end{figure}

The observed multiplicity distribution must be corrected for detector
effects (acceptance and efficiency in track reconstruction, 
spurious tracks from photon conversions and particle interactions 
in the material, selection cuts) and for effects induced by
the residual background.
In figure~\ref{mult1} we show the bias produced on the charged 
multiplicity distribution by the presence of residual \WW\ and
radiative events as well as the bias produced by selection cuts.
In figure~\ref{mult1}-a we compare two normalized multiplicity distributions
as they would appear in an ideal detector, namely without particle loss
and interactions in the material, after event selection.
In terms of real data, they would correspond to detector level corrected
distributions including residual initial state radiation (i.s.r.). 
%
%         Bias produced by WW bkg., i.s.r., selection cuts  
%
The dotted distribution is relative to the pure \qq\ sample which 
survived cuts, the other distribution (histogram) contains also
the residual contamination from \WW\ events.
As it can be seen from the bin-by-bin ratio shown in the bottom part,
the bias due to this kind of background is relatively small.
The estimated effect is $1.5\%$ on $<n_{ch}>$ and $0.6\%$ on $R_2$. 
In figure~\ref{mult1}-b a similar comparison is done to estimate the 
bias due to the presence of radiative events.
Although this kind of contamination is relevant, the difference 
between the distribution containing residual radiative \qq\
events (histogram) and the corresponding distribution for 
non-radiative events alone (dotted), is marginal.
The effect is negligible on $R_2$, while on $<n_{ch}>$ is similar
in size and in the opposite direction with respect to the one produced by 
\WW\ events. 
The bias introduced by selection cuts is important at high multiplicity.
This can be seen in figure~\ref{mult1}-c where the normalized 
distribution for non-radiative \qq\ events surviving the selection 
criteria, (histogram), and the normalized distribution for an 
unbiased \qq\ sample, (dots), are compared.
In this case the effect on $<n_{ch}>$ and $R_2$ was estimated to
be $3.7\%$ and $1.3\%$, respectively. 

Detector dependent corrections are usually carried out with  
unfolding matrix procedures and bin-by-bin coefficients\cite{matrix}.
In general matrices and coefficients are computed with the help of 
a very detailed detector Monte Carlo simulation in terms of material 
distribution, physical processes which particles undergo when 
interacting in the material and detector response to particles 
traversing the active media.
After subtraction of the estimated residual \WW\ contamination, 
using for example a bin-by-bin correction, and
provided one has a reliable simulation of the initial state radiation process,
a global correction for particle loss due to detector effects is 
conceivable using a single unfolding matrix, computed from fully 
simulated events including i.s.r.
The bias produced by event selection and residual i.s.r. can be
corrected using bin-by-bin coefficients. 

\begin{figure}[htbp]
\vspace{0.1cm}
\centerline{
\epsfig{figure=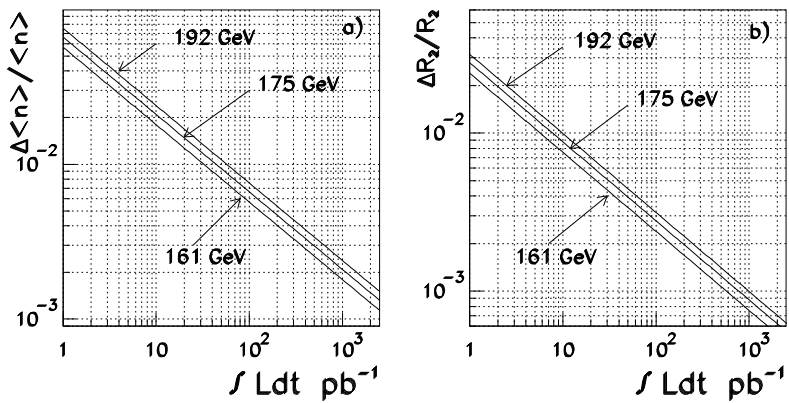,height=6cm,angle=0}
}
\caption{\it Expected relative statistical uncertainties on
$<n_{ch}>$ and $R_2$.}
\label{mult2}
\end{figure}

Considering our estimated selection efficiencies and assuming 
cross sections and multiplicity distributions as predicted by Pythia,
we show in figure~\ref{mult2}-a,b the expected relative statistical 
uncertainties on $<n_{ch}>$ and on $R_2$ as a function of the integrated
luminosity, at three different energies.
%
%      Statistical uncertainties on <n_ch> and R_2 vs Luminosity 
%
The integrated luminosities expected at \LepII\ are such
that statistical uncertainties on these parameters should comfortably 
stay below the $1\%$ level.
A sensible estimate of the magnitude of systematic uncertainties 
is difficult at this time.
It will be most probably dominated by model dependent corrections 
needed to handle residual background and event selection biases.
It is hard to believe it will be smaller than at \LepI\ ($1-2\%$ on
$<n_{ch}>$), but an uncertainty of a factor two higher
may not be out of reach.  
We have fit the average charged multiplicity measured above the 
Upsilon threshold\cite{comp} using the most popular 
parametrisations\cite{param}:
\[ \begin{array}{l}
   < n_{ch} > = a \cdot \alpha_{s}^\beta \cdot \exp(\gamma/\sqrt{\alpha_s}) \\
   < n_{ch} > = b \cdot s^a \\
   < n_{ch} > = a + b \cdot \ln s + c \cdot \ln^{2} s 
\end{array} \]
where a, b and c are free parameters.
The predicted values extrapolated to \LepII\ 
energies, however, only differ by a $3\%$ or less and it will be not 
obvious to disentangle among these models.

We also investigated the possibility to measure the average 
charged multiplicity in heavy-quark initiated events.
Vertex-tagging methods have been shown to be very 
effective to select b-quark samples of high purity,
and it is well known that secondary vertices with a
relatively high associated multiplicity are likely to 
be produced in this kind of events. 
In the present study a method recently applied at \LepI\cite{hfopal}
was used to analyse events simulated at $\sqrt{s}=175$~GeV.
%\hfill\break
%
%           Flavor composition vs decay length significance
%
\begin{figure}[htbp]
\vspace{0.1cm}
\centerline{
\epsfig{figure=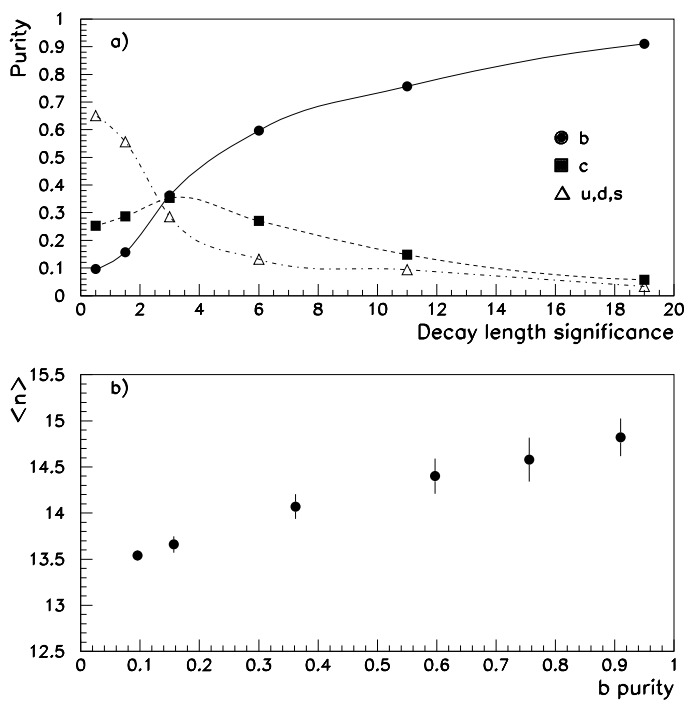,height=10cm,angle=0}
}
\caption{\it a) Flavour composition vs. decay length significance;
b) $<n_{ch}>$ for the unbiased event hemisphere vs. b-quark purity.}
\label{mult3}
\end{figure}
The method relies on the fact that independent samples of events with
a different flavour composition can be selected by requiring events 
to have (at least) one secondary vertex with a certain decay length 
significance, defined as the decay length divided by its error. 
The fraction of a given flavour in the sample can be evaluated, 
as a function of this variable, from fully simulated and reconstructed 
events, figure~\ref{mult3}-a.
In general to a high value of the decay length significance corresponds
a high probability to tag a b-quark event while at low significances
the samples are predominantly populated by light-flavour events.
To minimise the bias on multiplicity introduced by vertex tagging 
requirements, each event is divided in two hemispheres by
a plane perpendicular to the thrust axis and the multiplicity is
measured only in the hemisphere opposite to the one containing a
secondary vertex, (unbiased hemisphere).
The average charged multiplicity for pure samples of b-, c- and 
light-quark events is computed from a simultaneous fit to the corrected 
average multiplicity of samples selected with different decay length 
significance, i.e. with different flavour composition, 
figure~\ref{mult3}-b. 
More details about the experimental procedure can be found in\cite{hfopal}.

Due to extra selection requirements needed to insure the presence of 
secondary vertices, the original sample is reduced by a $30\%$. 
A b-tagging efficiency higher than $20\%$ can be achieved in the highest 
purity bin.
These values are similar to those obtained at \LepI.
We studied the effects induced by the residual background and by the
event selection cuts on the average charged multiplicity, measured in 
the unbiased hemisphere, as a function of the b-purity.
Again we find that \WW\ and radiative events produce only a 
marginal effect while the bias produced by selection cuts is important.
The unfolding matrices to correct for detector acceptance, efficiency and
spurious tracks must be calculated for each bin of decay length significance.
A bin-by-bin correction method could be used to unfold residual i.s.r. and
selection cuts effects.

In order to estimate the statistical precision attainable in this
kind of measurement, we used the selection efficiencies 
found in this study and assumed the cross sections
as well as the average multiplicity for different quark flavours,
$<n_q>$, predicted by Pythia.   
The fitting procedure mentioned above was applied to 
a high statistics sample of unbiased \qq\ events
to estimate the uncertainty on $<n_q>$. 
 
\begin{figure}[htbp]
\vspace{0.1cm}
\centerline{
\epsfig{figure=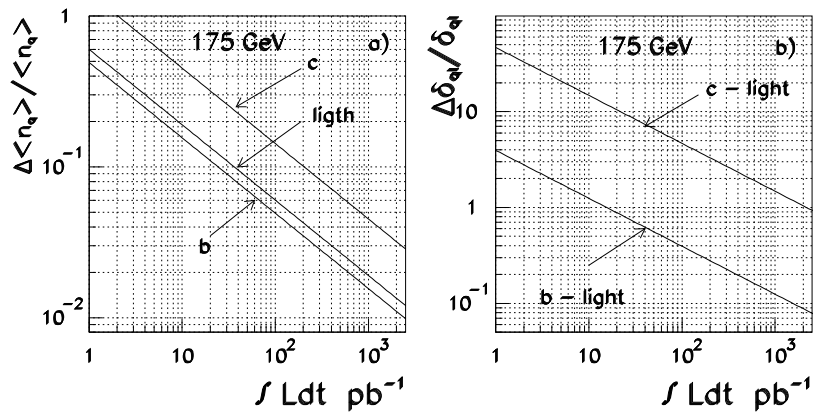,height=6cm,angle=0}
}
\caption{\it Expected relative statistical uncertainties on $<n_q>$
and $\delta_{ql}$.}
\label{mult4}
\end{figure}

In figure~\ref{mult4}-a we show the expected relative statistical 
uncertainties on $<n_q>$ (q = b,c,light) as a function of the 
integrated luminosity.
%
%     Expected statistical uncertainties on <n_q> 
%
The difference in charged multiplicity between
b- and light-quark events, $\delta_{bl}$, and between c- and light-quark
events, $\delta_{cl}$, is shown, taking into account 
correlations, in figure~\ref{mult4}-b.
One can see that these measurements will be largely dominated 
by statistical uncertainties, at least using this method of analysis.
A measurement of $\delta_{bl}$ could probably be attempted,
while a determination of $\delta_{cl}$ seems to be precluded. 

\section{Hadron Momentum Spectra as a Test of LLA QCD\protect\footnote{
Author: L.A. del Pozo}}
\label{DelPozo}
The shape of the momentum spectrum of hadrons produced in \epem\
collisions is successfully predicted in
leading-log QCD (LLA). The LLA family  of calculations   together
with the assumption of local parton hadron duality (LPHD)
\cite{LPHD} predict that
soft gluons should interfere destructively due to their coherent
emission (or angular ordering), and this gives rise to a `hump-backed'
shape for the momentum distribution. At leading order, the
distribution of $\ln(1/x)$ should have a Gaussian form, and this shape
is modified to be a Gaussian with higher  moments if
next-to-leading order terms are calculated. The position of the peak
of the distribution  $\ln(1/x_0)$ is predicted in terms of the centre-of-mass
energy, and therefore the  evolution of the peak position with energy
is well defined. 
This has been measured at centre-of-mass energies between 14 and
91~GeV, and the results are in agreement with a theoretical prediction
that includes the effects of coherent, soft gluons. Models for hadron
production based on phase space alone, or incoherent parton branchings
 predict a peak variation with energy that is twice as rapid and which is
 not supported by the data \cite{schmelling}\cite{thesis}. 
%{\em add some references here. xxx}

The increase of centre-of-mass energy afforded by the energy upgrade LEP2,
allows the hadron $\ln(1/x)$ distribution to be measured in  a new
energy regime and provides the opportunity to further test the
evolution of $\ln(1/x_0)$  from low energies. The energy increase is
of the order of a factor two, so this represents a substantial `lever
arm' when compared to the existing data. In order to be able to
challenge the predictions, the peak $\ln(1/x_0)$ should be measured
to  a precision of less than about 0.1 unit of $\ln(1/x_0)$. 

%{\em Possibly add a figure of the evolution of $\ln(1/x_0)$ with E..??}

The detailed shape of the  $\ln(1/x)$ is predicted in terms of a small
(typically three) number of parameters, and an energy evolution. These
parameters have been fixed by fitting to the LEP1 data
\cite{OP_1x}\cite{thesis},
and they can be used to predict the form of the data at higher energies.
Clearly such a prediction of the shape of the  $\ln(1/x)$ distribution
constitutes an important potential measure  of the  success of
the LLA  approach to QCD calculations.

The LLA approach has been extended to predict the momentum
distribution of pairs of gluons which has 
commonly been presented in terms of
the two particle correlation \cite{TPC_theory}. This distribution has
been measured at LEP1 \cite{thesis}\cite{TPC_paper} where it was
found that the 
data were qualitatively, but not quantitatively described by
next-to-leading order predictions. It was subsequently shown
\cite{dallas} that a
satisfactory description of the data was possible if
next-to-next-to-leading order terms with coefficients of order unity
were added to the prediction.
The study of the energy dependence of the two particle correlation is 
interesting as it is predicted solely in terms of a single free
parameter and the energy scale.

\subsection{Monte Carlo Studies at 175 GeV}

Monte Carlo events generated at a centre-of-mass energy of 175~GeV
were used to study the likely precision and limitations of an analysis of
hadron momentum spectra at LEP2. Events were generated using Pythia
for the processes 
${\mathrm e}^+{\mathrm e}^- \rightarrow {\mathrm W}^+{\mathrm W}^- 
\rightarrow {\mathrm q}\overline{{\mathrm  q}}
{\mathrm q}\overline{{\mathrm  q}}$
  and   ${\mathrm e}^+{\mathrm e}^- \rightarrow {\mathrm Z}^0/\gamma
\rightarrow {\mathrm q}\overline{{\mathrm  q}}$,
and were subsequently passed through the Opal detector simulation
program. 

In contrast to analyses at LEP1 energies, the chief experimental
problems are  event statistics, and backgrounds in the event sample
due to ${\mathrm W}^+{\mathrm W}^- 
\rightarrow {\mathrm q}\overline{{\mathrm  q}}
{\mathrm q}\overline{{\mathrm  q}}$ events and events with a large
amount of energy radiated by the initial fermions.
The efficient selection of a clean sample of $\rightarrow {\mathrm Z}^0/\gamma
\rightarrow {\mathrm q}\overline{{\mathrm  q}}$ events with propagator
energies close to the centre-of-mass energy has been extensively
studied in subsection~\ref{sect-Zsel}. The present 
 analysis uses the stage I cuts
described there  together with a stage II cut of 
$D^2 > 300  {\mathrm G}{\mathrm e}{\mathrm V}^2$ and $B_N<0.05$. In
total about 
9000 ${\mathrm e}^+{\mathrm e}^- \rightarrow {\mathrm Z}^0/\gamma$
events and about 130 ${\mathrm e}^+{\mathrm e}^- \rightarrow {\mathrm
  W}^+{\mathrm W}^- $ events would be  selected assuming the nominal
LEP2 luminosity of 500 \mbox{pb$^{-1}$} and standard model cross
sections.  About 72~\% of the selected
events have initial state radiation amounting to  less than 2~GeV, and
less than 2~\% of the events have radiation in excess of 60 GeV.

\subsection{ln (1/{\em x}) Distributions  at 175 GeV}

\begin{figure}[th]
\begin{center}
\mbox{
\epsfig{%
file=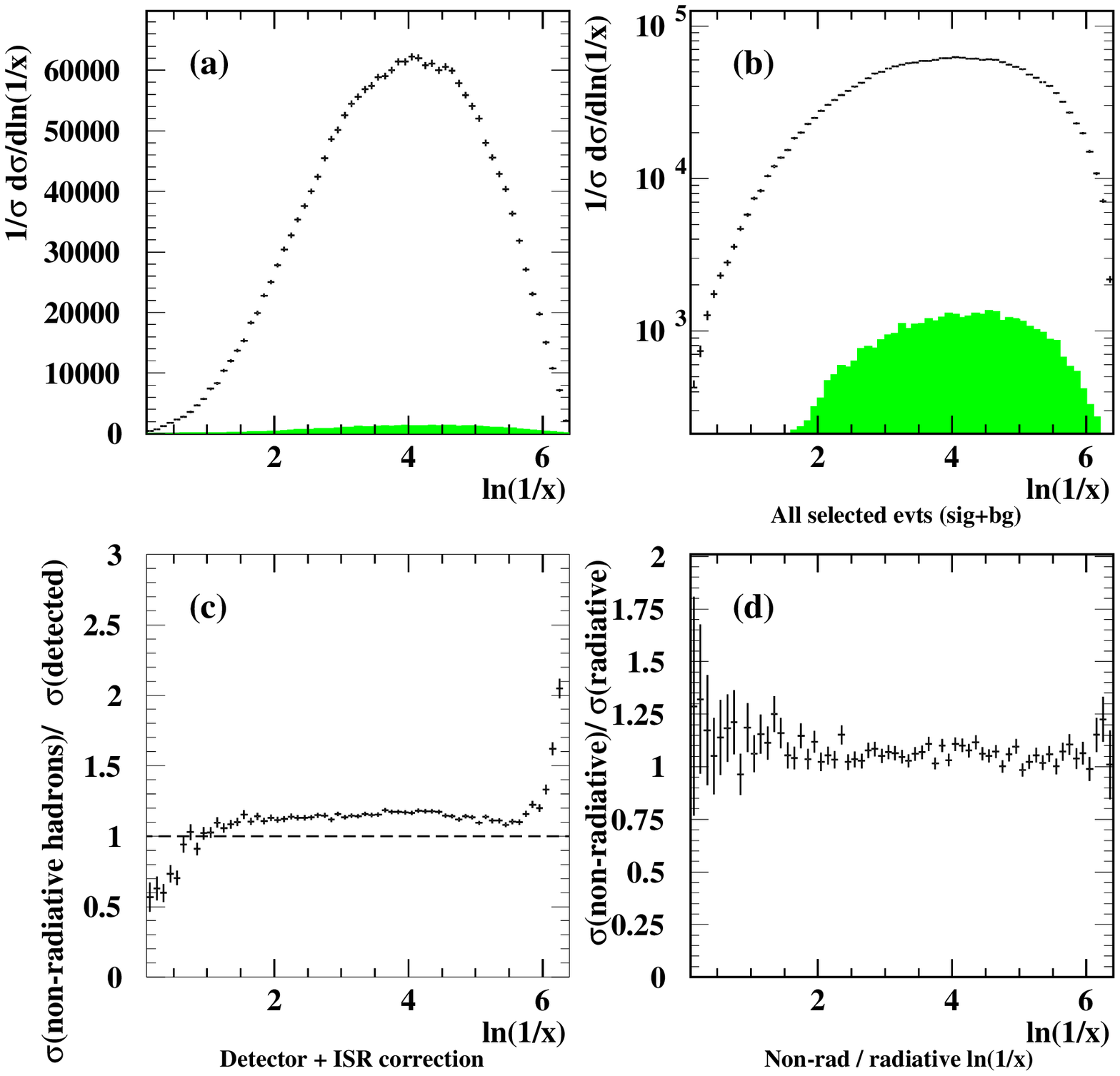,width=0.9\textwidth,%
bbllx=31pt,bblly=170pt,bburx=536pt,bbury=657pt
}}
\caption{Distributions of $\ln(1/x)$ for 500 pb$^{-1}$ of events
  passing selection cuts. Background from  $ {\mathrm  W}^+{\mathrm
    W}^-$ events is shown as the shaded areas. Figures (c) and (d)
  show the the typical acceptance and initial state radiation
  corrections that might be expected.}
\label{fig_1x}
\end{center}
\end{figure}

The expected distribution of $\ln(1/x)$ is shown in
figure~\ref{fig_1x}~(a) (and figure~\ref{fig_1x}~(b) with a
logarithmic vertical scale)  for all events that pass the selection cuts.  The
statistical errors on 
the points correspond to a luminosity of 500~pb$^{-1}$. The
contribution of the 
 ${\mathrm W}^+{\mathrm W}^-$ background events is shown as the shaded
 region. The background is concentrated in the region 
around the peak  of the  $\ln(1/x)$ distribution, varies smoothly, and is small
compared to the level of the signal events -- the signal to background
ratio is almost 100:1. In practice, this background could be corrected
for by a multiplicative correction factor, the background could be
subtracted directly or a more complex matrix correction procedure 
could be applied.

Typical  corrections for detector acceptance
and resolution are shown in figure~\ref{fig_1x}~(c).
 There is a correction of about 10~\% to the overall
level of the distribution which is basically flat in the region  around
the peak of the $\ln(1/x)$
distribution. Uncertainties in the determination of  detector 
corrections are therefore 
unlikely to have a large effect on the position of the
peak, and there is no evidence that  a serious bias  has been
introduced into the $\ln(1/x)$ distribution by the event
selection. Figure~\ref{fig_1x}~(d) shows the ratio of
the $\ln(1/x)$ distributions for events passing the selection cuts
that did not radiate and those that radiated a photon of more than
2~GeV. This illustrates the component of the detector correction that
accounts for  initial state radiation. The bias introduced by
 the initial state radiation  is most severe for  low values of $\ln(1/x)$
  but is fairly uniform around the area of the peak. It is
 not expected that the event selection and detector corrections will
 seriously bias the measurement of the position of the peak.

\subsection{Determination of Peak Position}
The peak position may be determined by fitting a Gaussian to the $ {\mathrm
  W}^+{\mathrm W}^-$-background subtracted
$\ln(1/x)$ distribution. The statistical error on the peak position is
about 0.025 if data corresponding to 100 pb$^{-1}$ are fitted, and this
decreases to 0.011 when  the expected 500 pb$^{-1}$ data sample is analysed. 
A Gaussian function is only valid for the region close to the
peak and is less successful at describing the shape of the
distribution far away from it. This leads to a variation of the fitted
peak position as data points far from the peak are included in the
fit. Varying the fit range such that a reasonable $\chi^2$ is still
obtained for the fit results in an uncertainty in the peak position of
about 0.02. 

A systematic error due to uncertainties in the level of the $ {\mathrm
  W}^+{\mathrm W}^- $ backgrounds has been estimated by varying the
amount of the background subtracted by $\pm 100~\%$. The fitted peak
position changes by less then 0.01 in all cases. It is not expected
that there is a large uncertainty due to the details of the shape of
the $ {\mathrm  W}^+{\mathrm W}^-$ background. The process $ {\mathrm
  W}^\pm \rightarrow {\mathrm q}\overline{{\mathrm  q}}$ is very
closely related to $ {\mathrm
 Z}^0 \rightarrow {\mathrm q}\overline{{\mathrm  q}}$  which has been
 very well understood thanks the the LEP1 data. 
In conclusion, it is expected that the position of the peak of the
$\ln(1/x)$ distribution may be measured with the data recorded at LEP2
to a sufficient precision in order to be able to test the LLA
predictions.

\subsection{Detailed Shape of  ln(1/{\em x}) Distribution}

The expected statistical errors on the points of the $\ln(1/x)$
distribution are small compared to the bin-to-bin variations of the
distribution -- there is no apparent scatter of the data points. This
indicates that the data will most likely be of sufficient precision to
allow a detailed comparison with the shape predicted by theoretical
calculations with parameters fitted to  LEP1 data. The comments
regarding the influence of acceptance, initial state radiation and  $
{\mathrm  W}^+{\mathrm W}^-$ background corrections on the peak
position also apply to the shape of the distribution -- they are not
expected to pose a major problem. If these systematic effects
 turn out to be troublesome, there is the  still the  prospect that a
reasonable measurement of the ratio of $\ln(1/x)$ distributions at LEP1
and LEP2 energies might  be made in which many systematic
effects may cancel. 

It might be expected that the description of data by LLA
predictions is more successful at higher energies as the LPHD
assumption is more justified. This is supported in part by Monte Carlo
studies that indicate that the differences between hadrons and partons
are much reduced at LEP2 energies. 

\subsection{Two Particle Correlation}

The two particle correlation at LEP2 energies 
has been studied in the same way as the
single particle $\ln(1/x)$ distribution. If the correlation
distribution is computed along lines in the $\ln(1/x_1)-\ln(1/x_2)$ plane
as in reference~\cite{TPC_paper} then the statistical error on each
point would be of the order of 0.02 for the full 500 pb$^{-1}$ data
ample.
This  should be compared to 0.005
achieved in reference~\cite{TPC_paper}  with about 21 pb$^{-1}$ of LEP1
data. Preliminary studies indicate that corrections for acceptance, 
resolution  and initial state radiation will be small as anticipated  for
this  distribution. As for the LEP1 analysis, it
 is also expected that other  systematic effects
such as the ${\mathrm  W}^+{\mathrm W}^-$ background might  also
cancel when the normalized correlation distribution is calculated.

With the luminosity currently expected from LEP2 it is expected that
any measurement of the two particle correlation would be statistics
limited. There is however the hope that if the entire data sample is
analysed, the possibility exists to test the energy evolution of the
predicted correlation distribution in a meaningful way. In particular
it can be tested whether the distribution at higher energies may be
fitted by a prediction with the coefficients of the
next-to-next-to-leading order  terms fitted to LEP1 data. Such a
prediction with coefficients fitted to LEP1 data is able to describe
Pythia/Jetset events at both 91 and 175~GeV. Finally, the ratio of the
two particle correlation at 91 and 175~GeV may be measured and
compared to the theoretical prediction with the advantage
that uncomputed higher order terms may cancel to some extent in the ratio.

\subsection{Summary}

In summary, measurements of hadron momentum spectra offer the
possibility to make detailed tests of LLA QCD predictions,
particularly in terms of their energy evolution. The peak position of the
$\ln(1/x)$ distribution may be measured accurately with only a small
amount of data allowing a powerful test of the extrapolation from
lower energies. It should also be possible to determine the detailed
shape of this distribution which  will provide a stringent test of 
the energy evolution of predictions from LEP1 energies. 
Meaningful measurements of the two particle correlation will probably
have to wait for the full 500 pb$^{-1}$ of luminosity to be delivered
by LEP2.

\end{document}